\documentclass[fleqn,usenatbib]{mnras}

\def\3x2pt{3$\times$2pt}
\def\5x2pt{5$\times$2pt}
\def\6x2pt{6$\times$2pt}

\usepackage{slashed}
\usepackage{graphicx}
\usepackage{grffile}
\usepackage[normalem]{ulem}
\usepackage{dcolumn}
\usepackage{xcolor}
\usepackage{booktabs}
\usepackage{amsmath}
\usepackage{amssymb}
\usepackage{slashed}
\usepackage{todonotes}
\usepackage{comment}

\usepackage{multirow}
\usepackage{arydshln}
\usepackage[T1]{fontenc}
\usepackage{ae,aecompl}
\usepackage{lineno}

\usepackage{txfonts}

\title[Cosmological Constraints from Lensing Ratios]{\boldmath Cosmological lensing ratios with DES Y1, SPT and {\it Planck}}

\date{\today}

\pubyear{2018}

\author[Prat, Baxter et al.]{
\parbox{\textwidth}{
{\fontsize{11.3}{12} \selectfont 
J.~Prat$^{1}$\thanks{E-mail: jprat@ifae.es},
E.~Baxter$^{2}$\thanks{E-mail: ebax@sas.upenn.edu},
T.~Shin$^{2}$,
C.~S{\'a}nchez$^{2}$,
C.~Chang$^{3}$,
B.~Jain$^{2}$,
R.~Miquel$^{4,1}$,
A.~Alarcon$^{5,6}$,
D.~Bacon$^{7}$,
G.~M.~Bernstein$^{2}$
% ...
\begin{center} (The DES and SPT Collaborations) \end{center}
}}
\vspace{0.4cm}
\\
\parbox{\textwidth}{Author affiliations are listed at the end of the paper}}

\usepackage{eso-pic}

\AddToShipoutPictureBG*{%
  \AtPageUpperLeft{%
    \hspace{0.75\paperwidth}%
    \raisebox{-3.5\baselineskip}{%
      \makebox[0pt][l]{\textnormal{DES-2018-0357}}
}}}%

\AddToShipoutPictureBG*{%
  \AtPageUpperLeft{%
    \hspace{0.75\paperwidth}%
    \raisebox{-4.5\baselineskip}{%
      \makebox[0pt][l]{\textnormal{FERMILAB-PUB-18-547-AE}}
}}}%

\begin{document}
\label{firstpage}
\pagerange{\pageref{firstpage}--\pageref{lastpage}}
\maketitle

\begin{abstract}
Correlations between tracers of the matter density field and gravitational lensing are sensitive to the evolution of the matter power spectrum and the expansion rate across cosmic time.  Appropriately defined {\it ratios} of such correlation functions, on the other hand, depend only on the angular diameter distances to the tracer objects and to the gravitational lensing source planes.  Because of their simple cosmological dependence, such ratios can exploit available signal-to-noise down to small angular scales, even where directly modeling the correlation functions is difficult.  We present a measurement of lensing ratios using galaxy position and lensing data from the Dark Energy Survey, and CMB lensing data from the South Pole Telescope and \textit{Planck}, obtaining the highest precision lensing ratio measurements to date.  Relative to the concordance $\Lambda$CDM model, we find a best fit lensing ratio amplitude of $A = 1.1 \pm 0.1$.  We use the ratio measurements to generate cosmological constraints, focusing on the curvature parameter. We demonstrate that photometrically selected galaxies can be used to measure lensing ratios, and argue that future lensing ratio measurements with data from a combination of LSST and Stage-4 CMB experiments can be used to place interesting cosmological constraints, even after considering the systematic uncertainties associated with photometric redshift and galaxy shear estimation. 
\end{abstract}

\section{Introduction}
\label{sec:intro}

As photons from a distant light source traverse the Universe, their paths are perturbed by the gravitational influence of large scale structure.  Since galaxies trace this structure, the projected galaxy density on the sky, $\delta_g$, is correlated with the strength of gravitational lensing, as quantified via the convergence, $\kappa$.  Two-point correlation functions between $\delta_g$ and $\kappa$ are sensitive to the cosmological growth of structure and to the geometry of the Universe \citep[e.g.][]{Bianchini:2015, Giannantonio:2016, Prat:2017}.
We refer to the galaxies used to compute $\delta_g$ as {\it tracer} galaxies since we use them as tracers of the large scale structure. 

Extracting useful cosmological information from tracer-lensing correlations is complicated by the need to model the relationship between the galaxy density field and the underlying matter field, i.e. galaxy bias \citep{Benson:2000}.  Furthermore, at small angular separations, lensing-galaxy two-point functions become sensitive to the small scale matter power spectrum, which is difficult to model due to e.g. nonlinearities and baryonic effects \citep{vanDaalen:2011, Takahashi:2012}.  For these reasons, many recent analyses \citep[e.g.][]{DESy1:2017} have restricted the usage of galaxy-lensing correlations to the regime where a simple linear bias model can be assumed and baryonic effects on the matter power spectrum can be neglected.  While this approach has the advantage of decreasing the complexity of the required modeling, it comes at the cost of increased statistical uncertainty.  

Several authors \citep[e.g.][]{Jain:2003,Bernstein2006,Hu:2007,Das:2009} have pointed out that if one considers suitably defined {\it ratios} between lensing-galaxy two-point functions, the dependence of these ratios on the galaxy-matter power spectrum cancels, but the ratio is still sensitive to the angular diameter distances to the tracer galaxies and to the sources of light used to measure lensing.  This sensitivity can be used to constrain cosmology via the distance-redshift relation. The cancellation of the galaxy-matter power spectrum is valid when two conditions are met: (1) the ratio is between two two-point functions that involve the same set of tracer galaxies, but sources at two different redshifts, and (2) the tracer galaxies are narrowly distributed in redshift.  

In principle, any two sources of light could be used to compute a lensing ratio.  However, as pointed out by \citet{Hu:2007} and \citet{Das:2009}, lensing ratios that involve galaxy light as one of the source planes and CMB light as the other are especially interesting cosmological probes.  There are two reasons for this.  First, the CMB provides a very long redshift lever arm, which increases the sensitivity of the ratios to cosmological parameters.  Second, the redshift of the CMB is known very precisely and is not subject to e.g. photometric redshift uncertainty.  In contrast, lensing ratios involving only galaxy lensing are more sensitive to photometric redshift and shear calibration errors, and less sensitive to cosmology because both source planes are then at low redshift.  Indeed, the recent galaxy-galaxy lensing analysis of \citet{Prat:2017} used lensing ratios to place constraints on the photometric redshifts of source galaxies, and demonstrated their ability to inform shear calibration priors as well. On the other hand, \citet{Kitching2015} used lensing ratio measurements involving only galaxy lensing with galaxy clusters as lenses to measure the distance-redshift relation, and infer cosmological parameters in combination with other probes. 

In this work, we present measurements of lensing ratios involving galaxy lensing and CMB lensing using data from the Dark Energy Survey (DES), the South Pole Telescope (SPT) and \textit{Planck}.  The DES data is used to construct samples of tracer galaxies and to generate weak lensing convergence maps. The SPT and \textit{Planck} data are used to construct CMB lensing convergence maps.  We measure angular correlations between the  tracer galaxy samples and the convergence maps, and use these measurements to constrain lensing ratios for multiple source and tracer galaxy redshift bins. The measured ratios are then used to constrain cosmology, focusing on the curvature parameter, $\Omega_{ k}$.  

For current data, with measurement uncertainty on the lensing ratios of roughly 10\%, the cosmological constraints obtained from the ratio measurements are fairly weak. We therefore also explore the potential of future data to constrain cosmology using lensing ratios.  In particular, we consider how the presence of systematic errors in estimated redshifts and shears can degrade the cosmological constraints from lensing ratio measurements. As part of this analysis, we consider how future lensing ratio constraints can potentially be improved by using photometrically identified tracer galaxies rather the spectroscopically identified galaxies, sacrificing some redshift precision for increased number density and increased overlap on the sky with planned CMB experiments.

An analysis of lensing ratios formed with galaxy lensing and CMB lensing measurements was recently presented by \citet{Miyatake:2017}. In addition to using different, more constraining data, the present work differs from that of \citet{Miyatake:2017} in two important respects. For the first time, we use a set of tracer galaxies obtained from a photometric survey. This is possible because of the redMaGiC algorithm \citep{Rozo2015}, which produces a selection of galaxies with tightly constrained photometric redshifts, whose error distributions are very well understood. Second, we perform a complete cosmological analysis to obtain parameter constraints from the lensing ratios.

Our measurements of the correlations between tracer galaxies and both galaxy and CMB lensing are similar to those of \citet{Baxter2016}.  However, in that work, the measured correlation functions were fit directly, rather than being used to compute lensing ratios.  The complications of galaxy bias and baryonic effects at small scales were circumvented by introducing additional freedom into the model for the small-scale galaxy-matter power spectrum.  The main advantage of the present work over \citet{Baxter2016} is the reduced complexity of the modeling and the fact that the constraints obtained here are purely geometrical in nature.  Similarly, forthcoming analyses from DES and SPT will perform a joint analysis of cross-correlations between DES data products and CMB lensing maps produced from a combination of SPT and {\it Planck} data (see \citealt{Baxter:2018} for an overview of the analysis and methodology and \citealt{5x2pty1results} for the results).  While such joint two-point analyses can place tight cosmological constraints, they are limited by our ability to model the data across a wide range of angular scales.

This paper is organized as follows. In Sec.~\ref{sec:formalism} we introduce the basic formalism for describing the lensing ratios; in Sec.~\ref{sec:data} we describe the data sets used in this work; in Sec.~\ref{sec:measurements} we describe the process of extracting constraints on the lensing ratios from the data; in Sec.~\ref{sec:model} we extend our modeling to include important systematic effects, and describe tests of the model's robustness; the results of our analysis of the data are presented in Sec.~\ref{sec:results}; we make forecasts for future experiments in Sec.~\ref{sec:projections}, with emphasis on the impact of systematic errors in measurement of source galaxy redshifts; we conclude in Sec.~\ref{sec:discussion}.

\section{Formalism}
\label{sec:formalism}
In this section we present the theory relevant to computing the lensing ratios of two point correlation functions between some set of tracer galaxies and gravitational lensing convergence, which can be reconstructed using either galaxy shear measurements at redshift of $z\sim 1$ or using the CMB at redshift of $z \sim 1100$. The lensing convergence, $\kappa$, in the direction $\hat{\theta}$ is given by
\begin{equation}
\label{eq:convergence}
\kappa(\hat{\theta}) = \frac{3}{2} \Omega_{\rm m} H_0^2 \int d\chi \, d_{A}^2(\chi) \frac{q_s(\chi)}{a(\chi)} \delta(\hat{\theta}, \chi), 
\end{equation}
where $\Omega_{\rm m}$ is the matter density parameter today, $H_0$ is the Hubble constant today, $\chi$ is comoving distance, $d_A(\chi)$ is the angular diameter distance to $\chi$, $a(\chi)$ is the scale factor, and $\delta(\hat{\theta},\chi)$ is the overdensity at a particular point along the line of sight.  We have defined the lensing weight function 
\begin{equation}
q_s(\chi) = \frac{1}{d_A(\chi)} \int_{\chi}^{\infty} d\chi' W_s(\chi') \frac{d_A(\chi,\chi')}{d_A(\chi')},
\end{equation}
where $W_s(\chi)$ is the normalized distribution of source light as a function of redshift and we use the notation $d_A(\chi, \chi')$ to represent the angular diameter distance between comoving distance $\chi$ and $\chi$'.  For the CMB source plane, the source distribution can be approximated as a Dirac $\delta$ function centered at the comoving distance to the surface of last scattering, $\chi^*$.  In this case, the lensing weight function becomes
\begin{equation}
\label{eq:lensingfunc}
q_{\rm CMB}(\chi) = \frac{d_A(\chi,\chi^{*})}{d_A(\chi^*)\, d_A(\chi)}.
\end{equation}

We are interested in correlations between $\kappa$ and the projected density of the tracer galaxies on the sky, $\delta_g(\hat{\theta})$.  For tracer galaxies whose normalized redshift distribution is described by $W_l(\chi)$, the projected density on the sky can be written as
\begin{equation}
\delta_g(\hat{\theta}) = \int d\chi W_l(\chi) \, \delta^{3D}_g(\hat{\theta},\chi),
\end{equation}
where $\delta^{3D}_g(\hat{\theta},\chi)$ is the 3D galaxy overdensity.

We write the two-point angular correlation between tracer galaxies and the lensing convergence as $w^{i\kappa}(\theta)$, where $i$ labels the redshift bin of the lenses and $\kappa$ can represent either the galaxy lensing map ($\kappa_s^j$ for the lensing map derived from the $j$th galaxy source bin) or the CMB lensing map ($\kappa_{\rm CMB}$). It is also useful to define the harmonic space cross-spectrum between the galaxy density and lensing fields, which we write as $C^{i\kappa}(\ell)$. Using the Limber and flat sky approximations, we have
\begin{equation}
\label{eq:cell}
C^{i\kappa}(\ell) = \frac{3}{2} \Omega_{\rm m} H_0^2 \int d\chi W_{l}(\chi) \frac{q(\chi)}{a(\chi)} b\left(\frac{\ell}{d_A(\chi)}, \chi \right) P_{\rm NL}\left(\frac{\ell}{d_A(\chi)}, \chi \right),
\end{equation}
where $q(\chi)$ is the lensing weight function corresponding to $\kappa$.  We have written the galaxy-matter power spectrum as a bias factor, $b(k, \chi)$, multiplied by the non-linear matter power spectrum, $P_{\rm NL}(k,\chi)$. We can now convert the harmonic space cross-correlation to the angular two-point function:
\begin{equation}
\label{eq:hankel}
w^{i\kappa}(\theta) =  \sum \frac{2\ell + 1}{4\pi} F(\ell)P_{\ell} (\cos(\theta)) C^{i\kappa} (\ell),
\end{equation}
where $P_{\ell}$ is the $\ell$th order Legendre polynomial and $F(\ell)$ describes additional filtering that is applied to the $\kappa$ maps.  

As described in \citet{Baxter:2018}, modes below $\ell < 30$ and above $\ell \gtrsim 3000$ in the CMB $\kappa$ maps generated by \citet{Omori2017} can be very noisy, or potentially biased.  We therefore filter the CMB maps to remove these modes.  Since we are interested in ratios between correlations with $\kappa_{\rm CMB}$ and with $\kappa_s$, we apply the same filter to $\kappa_s$ as we use for $\kappa_{\rm CMB}$; this ensures that the expectation of the ratio of the correlation functions remains a constant function of angular scale.  Following \citet{Baxter:2018}, we adopt the filter function
\begin{equation}
F(\ell) = \exp (-\ell(\ell + 1)/\ell_{\rm beam}^2) \Theta(\ell - 30) \Theta(3000 - \ell),
\end{equation}
where $\ell_{\rm beam} \equiv \sqrt{16 \ln 2}/\theta_{\rm FWHM} \approx 2120$ and $\Theta(\ell)$ is a step function.  The use of the Gaussian smoothing reduces ringing as a result of the low-pass filtering.

In the limit that the tracer galaxies are narrowly distributed in redshift, the $W(\chi)$ factor in Eq.~(\ref{eq:cell}) can be replaced by $W(\chi) = \delta(\chi - \chi_l)$, where $\chi_l$ is the comoving distance to the tracer galaxies. After transforming to an integral over redshift, the ratio of the galaxy-CMB lensing cross-correlation to the galaxy-galaxy lensing cross-correlation can then be expressed as 
\begin{equation}
\label{eq:ratio}
r^{ij} = \frac{w^{i\, \kappa_{\rm CMB}}(\theta)}{w^{i \kappa_s^j}(\theta)} = \frac{d_A(z_l^i,z^*)}{d_A(z^*) \int_{z_l^i}^{\infty} d z \, n^j_s(z) \frac{d_A(z_l^i, z)}{d_A(z)}},
\end{equation}
where $n_s^j(z)$ is the normalized redshift distribution of the source galaxies and $z_l^i$ is the redshift of the tracer galaxies in the $i$th bin.  Eq.~\ref{eq:ratio} depends only on the redshift to the tracer galaxies, the source galaxies, and the surface of last scattering. Therefore, the lensing ratios depend only on the distance-redshift relation in this limit.  This is the main selling point of  lensing ratios as cosmological observables: they contain information about the expansion history of the Universe, but do not require modeling galaxy bias or the matter power spectrum to extract this information.  \citet{Bernstein2006} has also pointed out that similar cosmographic measurements using a combination of gravitational lensing and observations of the transverse baryon acoustic oscillation feature can constrain curvature without assuming anything about the dynamics or content of the Universe.  This is in contrast to other cosmological observables --- including the angular scale of the CMB power spectrum and measurements of the distances and redshifts of supernovae --- for which the dynamics must be specified in order to translate constraints on the distance-redshift relation to a constraint on curvature.  In this work, however, we will specify the dynamics by considering models with dark energy parameterized by an equation of state $w$.

In the analysis presented here, the tracer galaxies have a non-zero extent in redshift, so the $\delta$-function approximation made above is questionable.  However, we will show in Sec.~\ref{subsec:thin-lens} that the width of the tracergalaxy redshift distribution is sufficiently narrow, and the error bars on the ratio measurements are sufficiently large, that the redshift distribution of the tracer galaxies can be approximated as infinitely {narrow}.  Additionally, the above model description assumes that all redshift and shear measurements are performed without biases.  In Sec.~\ref{sec:model} we will extend the model to include parameterizations for  systematic errors in the measurements. 

\section{Data}
\label{sec:data}

In this work, we measure correlations between the tracer galaxies and lensing convergence maps generated from both the CMB and source galaxies. We use data from the first year observations of DES \citep{Flaugher2015,DES2016,y1gold} for both the tracer galaxy sample and the galaxy lensing convergence maps \citep{des_mm_2017}. For the CMB convergence map, we use the map described in \citet{Omori2017}, which used a combination of CMB data from SPT and \textit{Planck}. Below we describe in more detail the tracer galaxies and the convergence maps used in this work. 

\begin{figure}
\begin{center}
\includegraphics[width=0.49\textwidth]{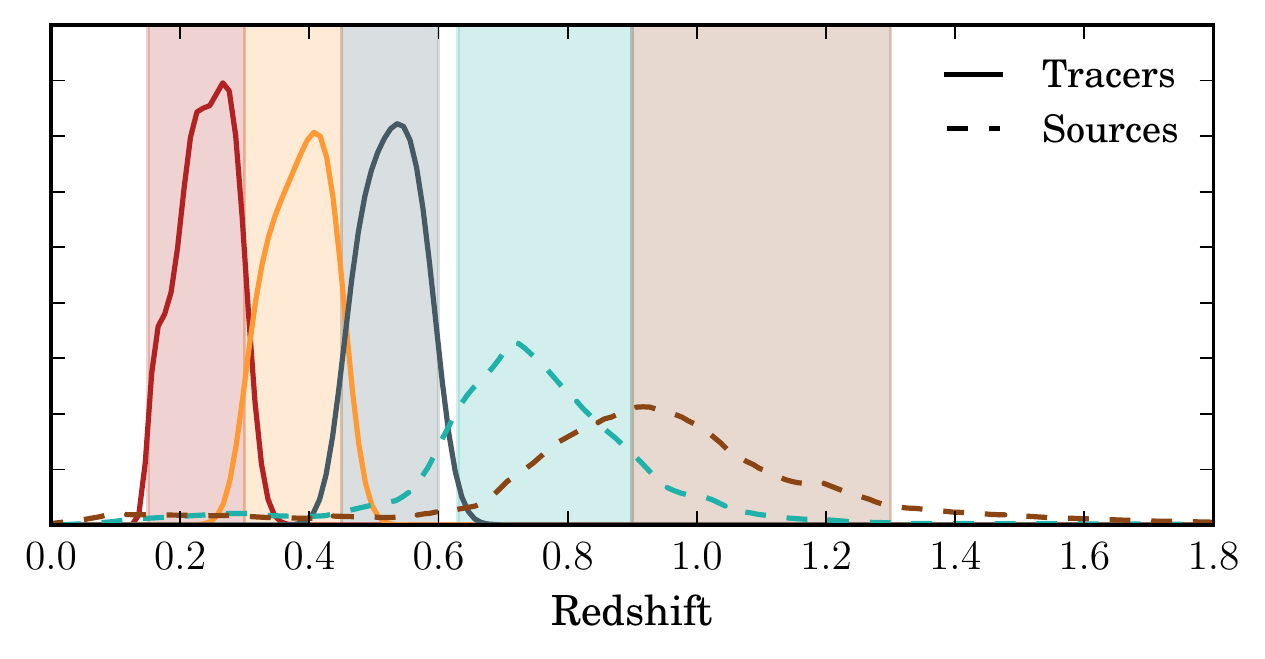}
\caption{The estimated redshift distributions of the tracer and source galaxies for the different bins used in this analysis.  Shaded bands represent the selection functions for the bins; galaxies are placed into bins according to the mean of their redshift probability distribution functions. }
\label{fig:nofzs}
\end{center}
\end{figure}

\subsection{Tracer galaxies}\label{subsec:lens-galaxies}

For the foreground tracer galaxies, we use a sample of galaxies referred to as ``redMaGiC.'' This sample is the same galaxy sample used in the DES Y1 cosmological analysis \citep{DESy1:2017}.
redMaGiC galaxies are luminous red galaxies selected based on goodness of fit to a red sequence template, as described in \citet{Rozo2015}.  The main advantage of the redMaGiC galaxy sample is that it is constructed to have very small photo-$z$ uncertainties. In particular, the DES Y1 redMaGiC photo-$z$s have a scatter of $\sigma_z = 0.0166(1+z)$. The tracer redshift distributions shown in Fig.~\ref{fig:nofzs} are computed from the sum of Gaussians with $\sigma = \sigma_z$, centered on the redshift estimates computed by redMaGiC for each galaxy. For a more detailed description of the tracer galaxy sample, see \citet{wthetapaper} and \citet{Prat:2017}.

We divide the tracer galaxies into three redshift bins between redshift 0.15 and 0.6, using the same $z$-binning as in the DES Y1 cosmology analysis \citep{DESy1:2017}.  The redshift distributions for these bins are estimated as the sum of the individual redshift probability distribution functions (PDF) for each redMaGiC galaxy, and are shown in Fig.~\ref{fig:nofzs}.  Galaxies were divided into bins based on the mean of the redshift PDF estimate for each galaxy. In this work we do not use the two higher redshift bins used by \citet{DESy1:2017} in order to minimize the overlap between tracers and sources. The uncertainty on the mean redshift for each of the redshift bins was studied in \citet{redmagicpz}, finding photometric redshift biases of $|\Delta \, z| < 0.01$.

\subsection{Galaxy lensing convergence maps}

We use the $\sim1300$ sq.~deg.~weak lensing convergence maps described in \citet{des_mm_2017}. These maps were generated from the DES Y1 \textsc{Metacalibration} shear catalog \citep{shearcat}, using the same sample that was used to obtain the DES Y1 3x2pt cosmology results \citep{DESy1:2017}. \textsc{Metacalibration} is a recently developed method to calibrate galaxy shear measurements from the data itself, measuring the response of a shear estimator to an artificially applied shear, without relying on calibration from simulations \citep{Sheldon2017,Huff2017}.  More details about the source sample and how the response corrections have been applied to the maps can be found in \citet{shearcorr} and in \citet{des_mm_2017}. 

The galaxy convergence maps of \citet{des_mm_2017} were constructed using an implementation of the Kaiser-Squires method \citep{Kaiser1993, Schneider1996} on a sphere \citep{Heavens2003,Castro2005,Heavens2006,Leistedt2017,Wallis2017}, which converts the shear, $\gamma$, into the convergence $\kappa$. The galaxy $\kappa$ maps used here were generated on \texttt{HEALpix} maps with $\texttt{nside}= 2048$, as opposed to the maps described in \citet{des_mm_2017}, which have $\texttt{nside}= 1024$. To match filtering applied to the CMB lensing maps described below, the galaxy $\kappa$ maps were smoothed with a 5.4 arcmin (FWHM) Gaussian, and were filtered to remove modes with $l < l_\mathrm{min} = 30$ and  $l > l_\mathrm{max} = 3000$.

We use the two higher redshift ($0.63 < z < 0.9$ and $0.9 < z < 1.3$) mass maps constructed in \citet{des_mm_2017} for this work.  The redshift distributions of the source galaxies used to construct these maps are shown in Fig.~\ref{fig:nofzs}, which have been obtained stacking a random sample from the redshift probability distribution of each galaxy. The source galaxy samples that were used to construct these two maps correspond to the two high-redshift source bins used in the DES Y1 cosmological analysis, and hence they have been studied extensively for both their photo-$z$ characteristics, \citep{Hoyle2018,Gatti2018,Davis2017,Prat:2017} and their shear measurement biases \citep{shearcat,y1-neighbours}. Briefly, their photometric redshift distributions have been estimated using the BPZ code \citep{Benitez2000a}, and calibrated using COSMOS galaxies and galaxy clustering cross-correlations with the redMaGiC sample. This allows us to use the results of these studies as priors in our model-fitting, which is essential for extracting cosmological information from the lensing ratios.

\subsection{CMB lensing map}

The CMB lensing map used in this analysis is presented in \citet{Omori2017}, and we refer readers to that work for more details.  Briefly, \citet{Omori2017} combined 150 GHz maps from SPT and 143 GHz maps from \textit{Planck} using inverse variance weighting to generate a combined CMB temperature map.  A quadratic lensing estimator \citep{Hu:2002} was then applied to the combined CMB temperature map to estimate $\kappa_{\rm CMB}$.  Bright point sources detected in the flux density range $50<F_{150}<500$ mJy and $F_{150}>500$ mJy in the 150 GHz band were masked with apertures of radius 6' and 9', respectively, prior to reconstruction.   Modes in the $\kappa_{\rm CMB}$ maps with $\ell < 30$ and $\ell > 3000$ were removed to reduce the impact of mean-field calibration and to suppress potential biases due to foregrounds, and a 5.4 arcmin Gaussian smoothing was applied, consistently with the galaxy $\kappa$ maps. 

We note that using the joint SPT+\textit{Planck} map from \citet{Omori2017} significantly improves the total signal-to-noise of the tracer-CMB lensing correlation measurements relative to using a CMB lensing map derived from {\it Planck} alone, and it also improves the results we would obtain with SPT alone.

\section{Measurements of the lensing ratios}
\label{sec:measurements}

In this section we describe the procedure for obtaining constraints on lensing ratios from the combination of DES, SPT and \textit{Planck} data.  We begin by describing the procedure used to measure the galaxy-lensing correlation functions and their corresponding covariance matrix. Next, we describe corrections for possible tSZ contamination of the CMB lensing maps.  Finally, we describe our fitting procedure for using the measured correlation functions to constrain the amplitudes of the lensing ratios.

\subsection{Measuring the tracer-lensing two-point functions}
\label{sec:measurement_procedure}

We measure the angular two-point correlation function between the pixelized lensing convergence maps $\kappa$ and the galaxy distribution $\delta_g$ by summing over tracer--convergence pixel pairs $g$, separated by angle $\theta$. We subtract the corresponding correlation with a sample of random points in place of the tracer galaxies, where the sum is over random--convergence pairs $r$ separated by $\theta$. The final estimator is
\begin{equation}
w^{i\kappa}(\theta) = \frac{\sum_{g} \omega_{g} \kappa_{g} }{\sum_{g} \omega_{g}} (\theta) - \frac{\sum_{r} \omega_{r} \kappa_{r} }{\sum_{r} \omega_{r}}(\theta),
\end{equation}
where $\omega_{g}$ and $\omega_{r}$ are the weights associated respectively with each tracer galaxy and random point.  This estimator is analogous to that used in tangential shear measurements in galaxy-galaxy lensing (see e.~g.~\citealt{Prat:2017}).  For the random points we set $\omega_{r} = 1$, and for the galaxies this weight was computed in \citet{wthetapaper} to reduce the correlation with observational systematics. For the fiducial measurements in this work, we grouped the tracer-convergence pairs in five log-spaced angular separation bins between 2.5 and 100 arcmin.  We use \texttt{TreeCorr} \footnote{$\texttt{https://github.com/rmjarvis/TreeCorr}$} \citep{Jarvis2004} to measure all two-point correlation functions in this work.  The measured correlation functions are shown in Fig.~\ref{fig:measurement}. 

\subsection{Covariance matrix of the two-point functions}

We estimate the covariance matrix between the measurements using the jackknife method.  In this approach, the survey area is divided into $N_{\mathrm{JK}}$ regions (`jackknife patches'), and the correlation function measurements are repeated once with each jackknife patch removed for the tracer sample, while we keep the convergence map untouched.  
The estimate of the covariance of measurements is then
\begin{equation}
C_{i\kappa \theta,i'\kappa'\theta'}^{\rm JK}= \frac{N_{\rm JK}-1}{N_{\rm JK}} \, \sum_{n=1}^{N_{\rm JK}} \, \left( \, w_{n}^{i \kappa} (\theta) - \overline{w^{i \kappa}}(\theta) \right) \, \left( \, w_{n}^{i' \kappa'}(\theta') - \overline{w^{i' \kappa'}}(\theta') \right),
\end{equation} 
where $i$ denotes the tracer galaxy bin, $\kappa$ denotes the convergence map, $n$ denotes the jackknife patch being removed, and $\overline{w^{i \kappa}}(\theta)$ is the mean across the $N_{\rm JK}$ resamplings.  The jackknife provides a {\it data}-based estimate of the covariance.  It is well motivated here since our analysis focuses on the small scales of the tracer-lensing correlations (down to 2.5') which are difficult to model theoretically.  Although the jackknife cannot capture super sample covariance \citep{Takada:2013} since by definition no samples are available outside the survey, this contribution to the covariance is expected to be negligible over the scales considered (i.e. below 100').  Moreover, at small scales, jackknife estimates have been extensively validated; see  e.g. \citet{Prat:2017} and \citet{NKpaper}.

The jackknife regions are obtained using the \texttt{kmeans} algorithm\footnote{\texttt{https://github.com/esheldon/kmeans\_radec}} run on a homogeneous random point catalog with the same survey geometry.  We choose $N_{\mathrm{JK}}=500$, which corresponds to jackknife regions whose typical size matches the maximum scale used in this work, of 100 arcmin.

\subsection{Correcting the two-point functions for thermal Sunyaev-Zel'dovich contamination}	
\label{sec:tsz_correction}

A study of the systematics affecting the 
$w^{i \kappa_{\rm{CMB}}}$
measurements using the DES redMaGiC galaxies and the \citet{Omori2017} CMB lensing map was performed in \citet{Baxter:2018}. In that work, the presence of the thermal Sunyaev-Zel'dovich (tSZ) effect in the CMB lensing map from \citet{Omori2017} was identified as a potentially significant source of contamination.  To reduce this contamination, \citet{Baxter:2018} took the conservative approach of excluding the small angular scales from the analysis that were estimated to be most contaminated.

Here, we take a more aggressive approach by explicitly modeling the tSZ contamination in our analysis.  We use the model of tSZ contamination from  \citet{Baxter:2018} for this purpose (see their Eq.~22).  The \citet{Baxter:2018} model was derived in the following manner.  First, the tSZ signal over the SPT patch was estimated using catalogs of galaxy clusters detected by DES, SPT and \textit{Planck}.  The tSZ signal for each cluster was estimated using a $\beta$-model \citep{Cavaliere:1976} fit to the observed cluster tSZ profile (for SPT-detected clusters) or by adopting a model profile given an estimate of the cluster mass (for DES and \textit{Planck} detected clusters).  The resulting tSZ map was then processed through the $\kappa$ estimation pipeline of \citet{Omori2017} to calculate spatially varying tSZ contamination in the $\kappa_{\rm CMB}$ maps.  Finally, the contaminant maps were correlated with the DES redMaGic catalogs to estimate the bias in $w^{i \kappa_{\rm CMB}}(\theta)$ due to tSZ contamination.  Fitting functions for the measured biases are provided in \citet{Baxter:2018}, and we adopt those fitting functions here.  

We test the sensitivity of our results to the model for tSZ contamination in Sec.~\ref{subsec:tsz_validation}.  Note that in this analysis, we make the same masking choices as in \citet{Baxter:2018} so that the tSZ model derived therein is appropriate; this includes masking the most massive galaxy clusters across the SPT field.  Note that in \citet{Baxter:2018} the effects of such masking on the correlation functions was found to be negligible relative to the statistical uncertainties.

\subsection{Extracting constraints on the lensing ratios}\label{subsec:extracting_constraints_ratios}

Given the measurements of the tracer-lensing correlation functions, we wish to extract constraints on the ratios of these correlations.  Simply taking the ratios of the correlation function measurements is not optimal when the two measurements have non-zero uncertainties and can lead to biased results.  Instead, we take the approach described below to measure the ratios. 

We model the correlation functions as
\begin{eqnarray}
 w^{i \kappa^j_{\rm s}} (\theta_a) &=& \beta_{ij} \alpha_{ia}  \label{eq:corr_model1} \\
     w^{i \kappa_{\rm CMB}} (\theta_a) &=& \beta_{i{\rm CMB}} \alpha_{ia}  f_i^{\rm tSZ}(\theta_a) \label{eq:corr_model2}
\end{eqnarray}
where $\alpha_{ia}$, $\beta_{ij}$ and $\beta_{i{\rm CMB}}$ are free parameters.  Here, $f^{tSZ}_{i}(\theta)$ is the tSZ bias model for each tracer bin $i$ described in Sec.~\ref{sec:tsz_correction}.  Without loss of generality, we set $\beta_{i0} = 1$.  In effect, the $\alpha_{ia}$ control the {\it shape} of the correlation function between the $i$th tracer redshift bin and each of the convergence maps.

On the other hand, the $\beta_{ij}$ and $\beta_{i{\rm CMB}}$ (which we can group as $\beta_{i \kappa}$), control the {\it amplitudes} of the correlation functions of different convergence maps with the tracer galaxies in redshift bin $i$; we will use the $\beta$s to extract constraints on the lensing ratios.  

Given our model for the measured correlation functions, we define a Gaussian likelihood for the measurements, $\{ w^{i \kappa} (\theta) \}$, where $\kappa$ can either be $\kappa_{\rm{CMB}}$ or the galaxy mass map in redshift bin $j$, $\kappa_s^j$:
\begin{multline}
\ln \mathcal{L}(\{ \, w^{i \kappa} (\theta) \, \} | \{ \alpha_i (\theta), \beta_{i\kappa}\}) =  
-\frac{1}{2} \sum_{i\kappa\theta \, i'\kappa'\theta'}  \left( w^{i \kappa} (\theta) -  \hat{w}^{i \kappa} (\theta)\right) \\
\times \left[\mathbf{C}^{-1}\right]_{i\kappa\theta,i'\kappa'\theta'}
\left( w^{i \kappa'} (\theta') - \hat{w}^{i \kappa'} (\theta')  \right). 
\end{multline}

In the equation above $\hat{w}$ represents the correlation function model from Eq.~\ref{eq:corr_model1} and Eq.~\ref{eq:corr_model2}, and $\mathbf{C}$ is the covariance matrix of the observations, as estimated with the jackknife method described in Sec.~\ref{sec:measurement_procedure}. We apply the so-called Hartlap factor \citep{Hartlap2007} to the inverted covariance to account for the noise in the jackknife covariance matrix estimate. We assume flat priors on the $\alpha$ and $\beta$, so the posterior on these parameters is simply proportional to the likelihood.  We sample from the model posterior using a Monte Carlo Markov Chain (MCMC) method implemented in the code \texttt{emcee} \citep{emcee}.

Ultimately, we are not interested in the $\alpha$ or $\beta$ themselves, but rather the ratios of the correlation function measurements for pairs that use the same tracer galaxy bin. We can obtain the posterior on the ratios by computing these ratios at each point in the Markov chains for the $\beta$'s.  At each point in the chains, we compute
\begin{eqnarray}\label{eq:ratio_betas}
 r_{ij} = \frac{\beta_{i\,\rm{CMB}}}{\beta_{i j}}.
\end{eqnarray}
The distribution of $r_{ij}$ then provides the posterior of the ratios, without loss of information.  By choosing to keep the (noisier) galaxy-CMB lensing two-point functions in the numerator of the ratio, we reduce the possibility of divergences in the ratios of the $\beta$s (which can occur if the posterior on a $\beta$ has support at $\beta =0$). Hereafter, we use the term \textit{lensing ratio} to refer to this definition of such ratios.

\section{Modeling the lensing ratios}
\label{sec:model}

Above we have developed a model for the correlation functions that allows us to extract constraints on the lensing ratios in Eq.~\ref{eq:ratio_betas}.   We now describe our parameterized model for the measured lensing ratios, including prescriptions for various systematic uncertainties, in order to extract constraints on cosmology. 

\subsection{Modeling photometric redshift and shear calibration bias}
\label{sec:sys_model}

As noted above, we assume that all of the tracer galaxies are located at a single redshift, $z_l$.  We obtain $z_l$ from the mean of the redshift distributions of the redMaGiC galaxies shown in Fig.~\ref{fig:nofzs}.  For the source galaxies, we use the full $n_s(z)$ when computing the model for the ratios. 

Following \citet{Krause:2017}, we parameterize redshift uncertainties in the estimated tracer and source galaxy redshift distributions with the bias parameters, $\Delta z_l$ and $\Delta z_s$, respectively.  This means that in Eq.~(\ref{eq:ratio}) we make the replacements 
\begin{equation}
n^j_s(z) \rightarrow n^j_s(z - \Delta z^j_s)
\end{equation}
and
\begin{equation}
\label{eq:photozbias}
z_l^i \rightarrow z_l^i + \Delta z_l^i,
\end{equation}
where $\Delta z_s^j$ and $\Delta z_l^i$ are treated as free parameters (with priors) for each source and tracer galaxy redshift bin, respectively.

We parameterize shear calibration bias with the parameter $m$ such that the observed shear is related to the true shear via $\gamma_{\rm obs} = (1+m)\, \gamma_{\rm true}$.  This means that we make the replacement 
\begin{equation}
\label{eq:shearbias}
r_{ij} \rightarrow \frac{r_{ij}}{1+m_j},
\end{equation}
where $m_j$ is a free parameter for each source galaxy redshift bin.

\subsection{Complete model for the lensing ratio}
\label{sec:complete_model}
Following from Eq.~(\ref{eq:ratio}) and including the above prescriptions for systematic uncertainties, our complete model for $r_{ij}$ is:  
\begin{eqnarray}
\label{eq:ratio_model}
\hat{r}_{ij}\, (\vec{\theta}_{\rm cosmo}, \vec{\theta}_{\rm sys} ) = \frac{(1+m_j)\,  d_A\, (z_l^i + \Delta z_l^i, z^*)}{d_A(z^*) \int_{z_l^i + \Delta z_l^i}^{\infty} dz \, n_s^j(z-\Delta z_s^j) \frac{d_A(z_l^i+\Delta z_l^i, z)}{d_A(z)}}, 
\end{eqnarray}
where $\vec{\theta}_{\rm cosmo}$ is the set of cosmological parameters and $\vec{\theta}_{\rm sys}$ is the vector of systematics parameters. We use \texttt{Astropy} for computing cosmological distances \citep{astropy2018}.

The posterior on the parameters given the set of measured ratios, $\{ r \}$, is given by
\begin{eqnarray}\label{eq:posterior-cosmopars}
P(\vec{\theta}_{\rm cosmo}, \vec{\theta}_{\rm sys} | \{ r \} ) = P( \{ r \} |\{\hat{r}(\vec{\theta}_{\rm cosmo}, \vec{\theta}_{\rm sys} )\})P_{\rm prior}(\vec{\theta}_{\rm cosmo})P_{\rm prior}(\vec{\theta}_{\rm sys}),
\end{eqnarray}
where $P_{\rm prior}(\vec{\theta}_{\rm cosmo})$ is the prior on the cosmological parameters, and $P_{\rm prior}(\vec{\theta}_{\rm sys})$ is the prior on the systematics parameters.  For the likelihood $P( \{ r \} |\{ \hat{r}\})$ we adopt a multivariate Gaussian approximation to the posterior from Sec.~\ref{subsec:extracting_constraints_ratios}:
\begin{eqnarray}
\ln  P( \{ r \} |\{ \hat{r}\})=-\frac{1}{2} (r - \hat{r}) \, \mathbf{C}_r^{-1} (r - \hat{r})^T.
\end{eqnarray}
We compute the covariance matrix of the ratio estimates, $\mathbf{C}_r$, from the Markov chains for the ratios described in  Sec.~\ref{subsec:extracting_constraints_ratios}.  We discuss the accuracy of the Gaussian approximation to the true posterior in Appendix~\ref{sec:gaussianity}. 

\begin{figure}
\includegraphics[width=0.5\textwidth]{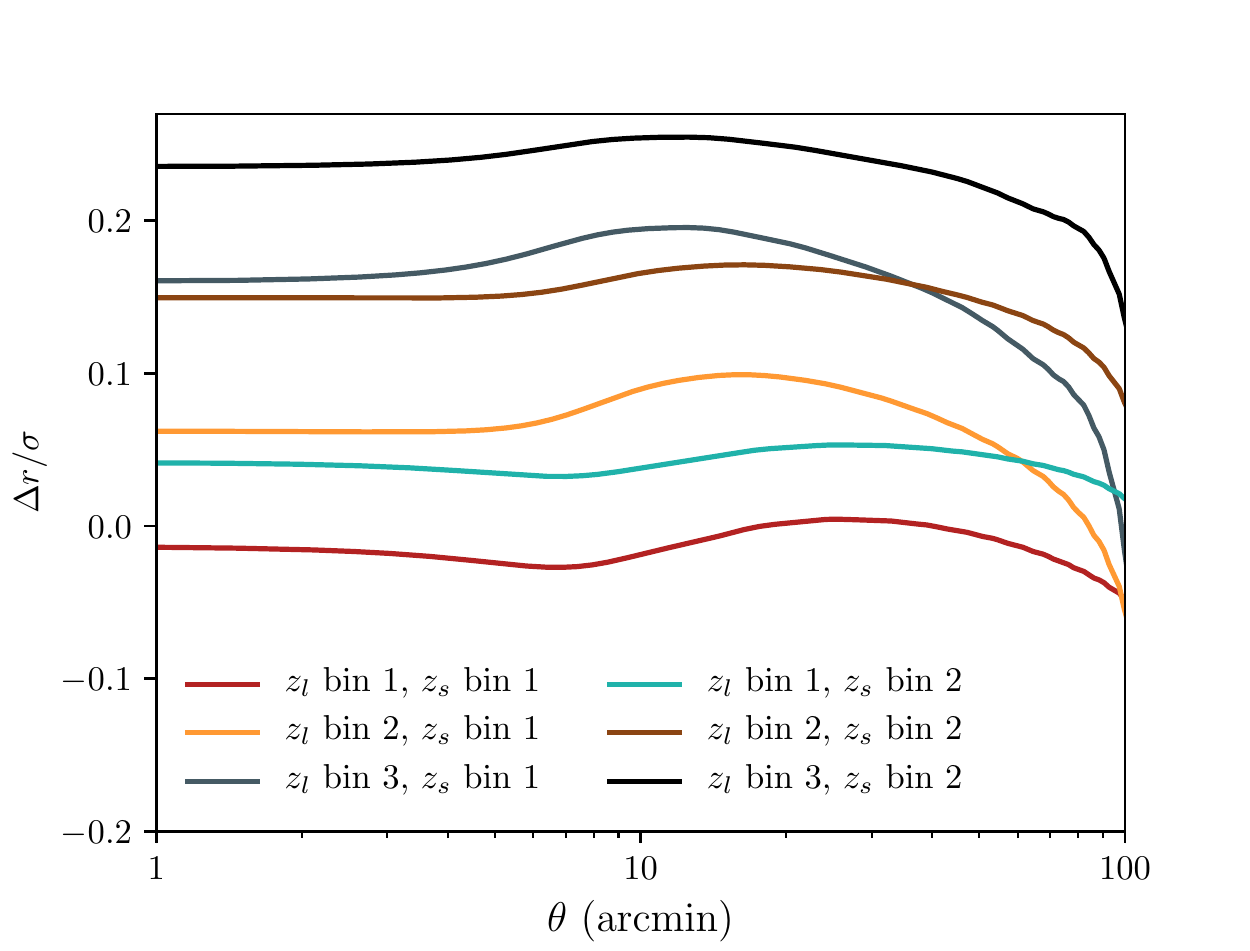}
    \caption{Test of the narrow tracer bin approximation used in this analysis.  We compute the error in the ratio, $\Delta r(\theta)$, incurred by assuming the tracer galaxies are distributed in infinitely narrow redshift bins.  This quantity is plotted relative to the statistical errors in the ratio measurements, $\sigma$, which is the uncertainty from all angular bins combined.  At most, the error incurred by assuming narrow tracer bins is $\sim 25\%$ of the statistical error on the ratio, and we therefore ignore it in this analysis.}
\label{fig:thin-lens}
\end{figure}

\subsection{Model Validation}
\label{sec:model_validation}

\subsubsection{Narrow tracer bin approximation}\label{subsec:thin-lens}

A fundamental assumption of our analysis is that in our modeling, we can approximate the tracer galaxy redshift distribution with a $\delta$ function centered at the mean of full redshift distribution.  Only in the $\delta$ function limit does the cancellation of the galaxy-matter power spectrum occur.  However, as described in Sec.~\ref{subsec:lens-galaxies}, the tracer galaxies used in this work are not all at the same redshift, but they are distributed over a relatively narrow redshift interval.  Furthermore, as seen in Fig.~\ref{fig:nofzs}, there is some overlap in the redshift distributions of the third tracer bin, and first source bin.  This overlap, which is not included in our modeling, will reduce the lensing signal (since sources at lower redshift than the tracers will not be lensed).

To test the impact of the narrow tracer bin approximation on our analysis, we generate simulated correlation function measurements using Eq.~(\ref{eq:hankel}).  To do this, we assume a linear bias model with $b = (1.45, 1.55, 1.65)$ for each of the tracer galaxy redshift bins, following the analysis of \citet{DESy1:2017}. Thus, for this test, we are assuming that the galaxy bias is independent of scale and of redshift within each tracer bin. The angular dependence of the ratio can then be computed from the simulated data vectors and compared to the approximate value of the ratio computed assuming infinitely narrow tracer redshift bins; we denote the difference between the true ratio and the approximated ratio as $\Delta r$.  

We plot the angular dependence of $\Delta r$ relative to the error bars on the ratio measurements in Fig.~\ref{fig:thin-lens}.  We see that for all tracer-source bin combinations, the error induced by the narrow tracer bin approximation is small compared to the error bars on the ratio.  Note that the decline in $\Delta r /\sigma$ close to 100 arcmin is due to the high-pass filtering that is applied to the lensing convergence maps.

\subsubsection{Lensing dilution and galaxy lensing boost factors}

\vspace{1mm}
When there is overlap in redshift between the source and the tracer galaxies two different effects occur. The first one is already mentioned in the section above, which is the dilution of the lensing signal when source galaxies are in front or at the same redshfit of tracer galaxies. In our analysis we make use of the narrow tracer bin approximation and therefore some of this dilution is not naturally accounted for in the theory prediction. Thus, to test for the impact of this effect, we have removed the bin combination which shows the largest overlap in redshift, which is the third tracer bin and first source bin combination (as seen in Fig.~\ref{fig:nofzs}), and found that removing it has negligible impact on the inferred cosmological parameters.

The second effect results from the tracer and source galaxies being physically correlated, since they both trace the large scale structure.  This changes the galaxy lensing signal since it will change the true $n(z)$ on the sky in a way that is not captured by the full survey $n(z)$.  Generally, this effect reduces the lensing signal since source galaxies behind the tracer galaxies will be on average closer to the tracer galaxies than what it is predicted by the full survey $n(z)$. To take into account this effect in the modelling we would need to measure the redshift distributions of the galaxies included in each of the angular bins. Alternatively, one can correct for this effect using the so-called boost factors. This correction is scale dependent and is bigger at small scales, where the clustering is also larger. Using the same data as employed here, \citet{Prat:2017} estimated the magnitude of this effect (i.e. the boost factor) by measuring the excess of sources around tracers compared to random points, as a function of scale, for every tracer-source bin combination (cf. their Figure 10). For the tracer-source binning configurations and for the choice of scales used in this analysis, the results in \citet{Prat:2017} demonstrate that the boost factors are 1\% or less over all angular scales, allowing us to safely ignore this effect in our analysis.  This makes sense, because we have attempted to use only tracer and source galaxy combinations that are well separated in redshift, so as to make the narrow tracer bin approximation more accurate.

\subsubsection{Intrinsic alignments}
\vspace{1mm}
Another systematic effect related to the overlap in redshift between the tracer and source galaxies is the intrinsic alignments (IA) of the shapes and orientations of source galaxies resulting from gravitational tidal fields during galaxy formation and evolution. IA can generate correlations between the source ellipticity and the lens position if the two galaxies are physically close. 

In Eq.~(\ref{eq:ratio}) we have assumed that there is no contribution from IA in the cross-correlation measurements with the galaxy convergence maps.  IA are detected in the multiprobe correlation function analysis of \citet{DESy1:2017}.  However, since here we analyze only those tracer and source redshift bin combinations that are widely separated in redshift, we expect the contribution from IA to be minimal for our analysis. Moreover, \citet{Blazek2012} found that when boost factors are not significant, IA can be ignored as well. 

\subsubsection{tSZ validation}\label{subsec:tsz_validation}
\vspace{1mm}
Our model of tSZ contamination of the measured two-point functions relies on estimating the tSZ signal for galaxy clusters across the SPT field.  To estimate possible systematic errors in our anlaysis associated with these modeling estimates, we recompute the bias corrections by modifying the assumed masses of the DES-detected clusters used when generating the contaminant maps.  The DES clusters dominate our estimate of the tSZ bias because the more massive SPT-detected clusters are masked.  The difference between the estimated biases for the fiducial and  perturbed models should therefore provide a reasonable estimate of our modeling uncertainty.  We generate two perturbed models by increasing and decreasing the amplitude of the assumed mass-richness relationship.  The fiducial mass-richness model is based on the weak lensing calibration of \citet{Melchior:2017}; the perturbed models adjust the amplitude of the normalization by $\pm 1\sigma$, where $\sigma$ represents the statistical uncertainty on the amplitude from the \citet{Melchior:2017} analysis.  Note that the updated weak lensing calibration of DES redMaPPer clusters by \citet{McClintock:2018} is consistent with that of \citet{Melchior:2017}, albeit with smaller error bars; using the $1\sigma$ error from \citet{Melchior:2017} is therefore a conservative choice.  We show the result of analyzing the data using our fiducial tSZ bias model and the two perturbed models in Sec.~\ref{sec:results}.  Note that simply varying the amplitude of the assumed mass-richness relation does not necessarily capture all of the uncertainty in the tSZ bias model.  However, since the tSZ amplitude scales strongly with mass, we expect the mass uncertainty to capture a dominant part of the total tSZ bias uncertainty.

\begin{figure*}
\centering
\includegraphics[width=0.7\textwidth]{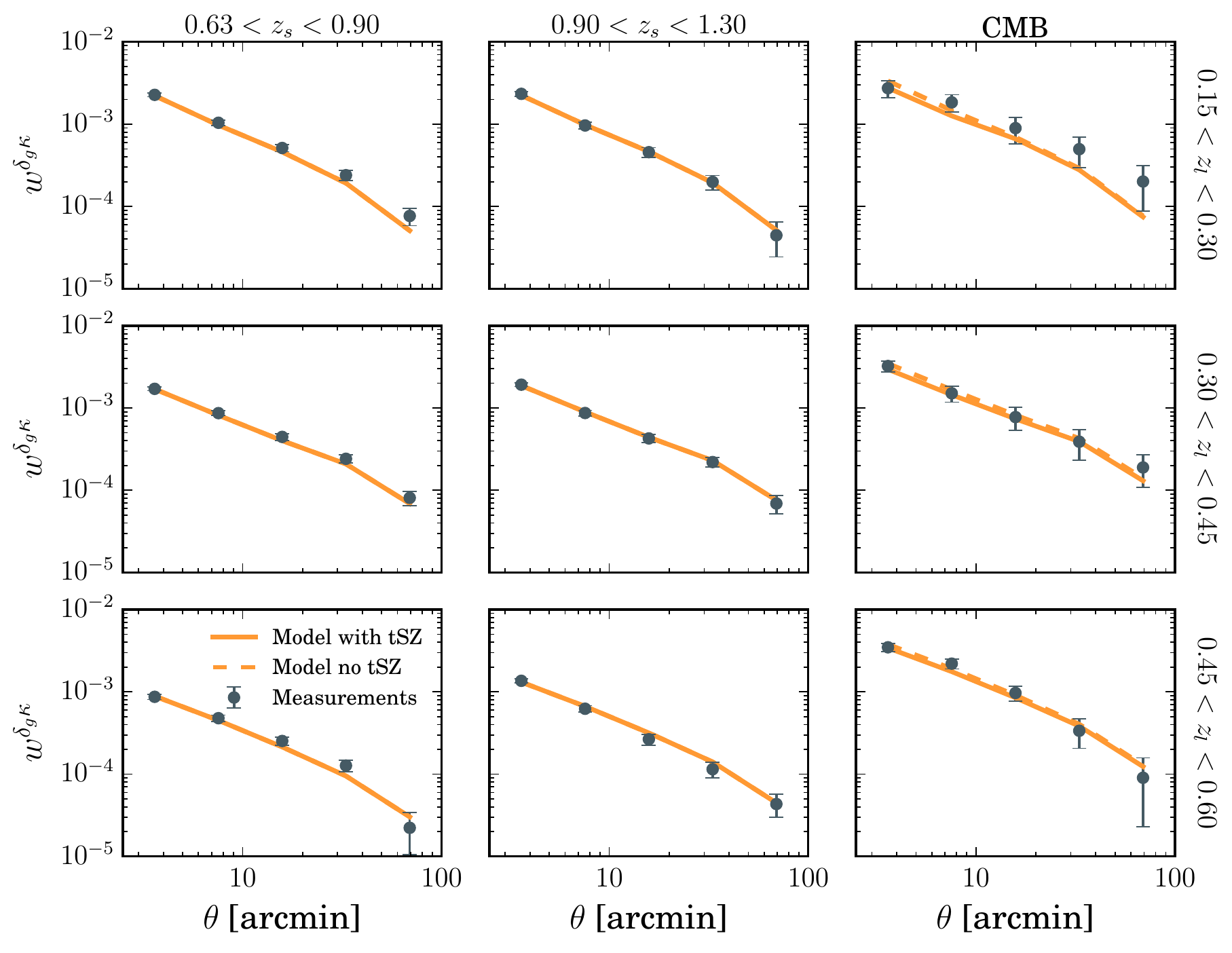}
    \caption{Tracer-lensing correlation function measurements, together with the best fit ratio model described in Sec.~\ref{subsec:extracting_constraints_ratios}. The model for the galaxy-CMB lensing correlations has been corrected for the tSZ-induced bias as explained in Sec.~\ref{sec:tsz_correction}. We also show the uncorrected model in dashed lines for comparison.}
\label{fig:measurement}
\end{figure*}

\begin{figure}
\centering
\includegraphics[width=0.5\textwidth]{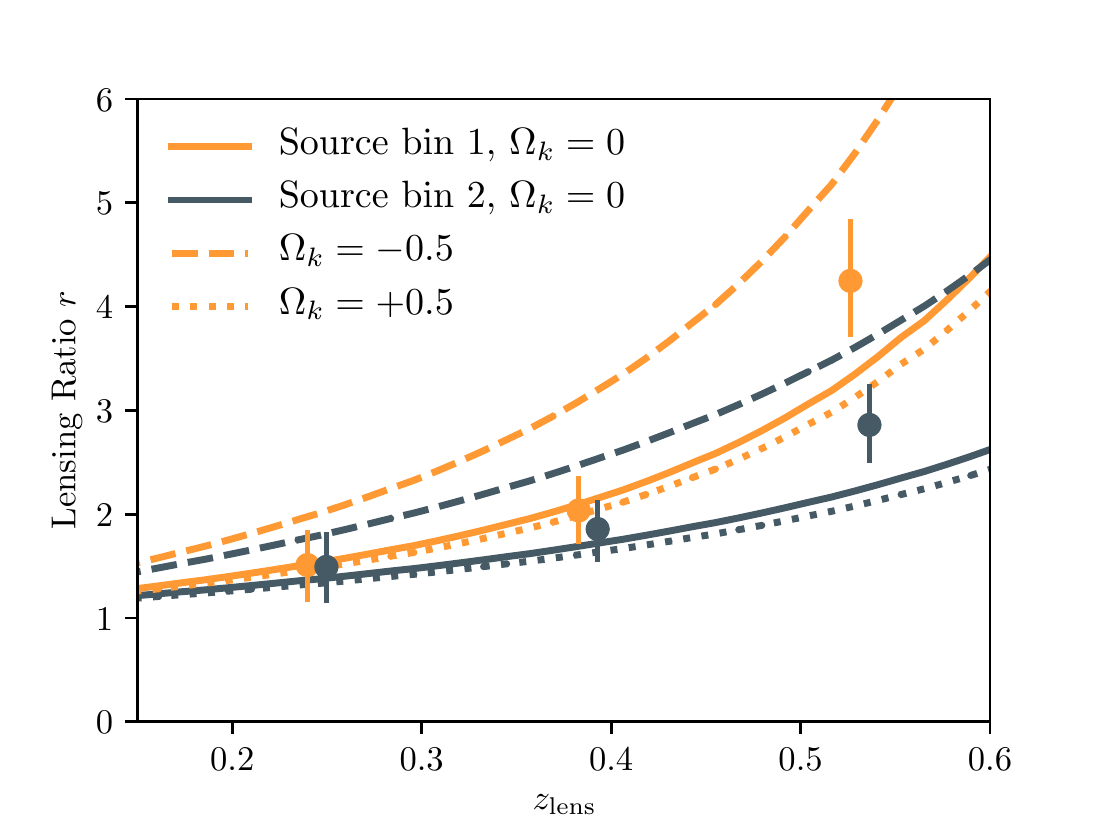}
    \caption{Measurements of the lensing ratios (points with error bars) as a function of lens redshift for two different source galaxy redshift bins (orange and gray curves).  The corresponding redshift distributions for these bins are shown in Fig.~\ref{fig:nofzs}.  Also shown are theoretical predictions (curves) for $\Lambda$CDM models with $\Omega_{\rm m} = 0.3$, but with different values of $\Omega_{ k}$.   Solid curves correspond to $\Omega_k = 0$, while the dashed and dotted curves change $\Omega_k$ to $-0.5$ and $0.5$, respectively.  Relative to the concordance flat $\Lambda$CDM model, the ratio measurements prefer an amplitude of $A = 1.1\pm0.1$, indicating consistency with this model.  While the highest redshift data points appear in some tension with the $\Omega_{k} = 0.0$ model, these points are covariant; the $\chi^2$ per degree of freedom relative to that model is $8.5/5$, corresponding to a probability to exceed of ${\rm p.t.e} = 0.13$.}
    \label{fig:miyatakestyle}
\end{figure}

\section{Results}
\label{sec:results}

We now present the constraints obtained on the lensing ratios and cosmological parameters from our analysis of  data from DES, SPT and \textit{Planck}.  We note that in order to avoid confirmation bias, our analysis was blinded during testing by replacing the true measurements with simulated data vectors.  The real data was used only after we were confident that the analysis pipelines were working correctly and the model had been validated. 

\subsection{Correlation function and ratio constraints}

The measurements of the two-point correlation functions between galaxies and (galaxy and CMB) lensing are shown as a function of angular scale in Fig.~\ref{fig:measurement} (\textit{points}), together with the best-fit ratio model described in Sec.~\ref{subsec:extracting_constraints_ratios} (\textit{lines}). For the cross-correlations with the $\kappa_{\rm CMB}$ map, 
we show both the model corrected 
by the tSZ effect (\textit{solid}), as described in Sec.~\ref{sec:tsz_correction}, and the uncorrected model (\textit{dashed}) for comparison.  

The corresponding constraints on the lensing ratios are shown in Fig.~\ref{fig:miyatakestyle} as a function of the mean lens redshift.  The full posteriors on the lensing ratios are shown in Fig.~\ref{fig:ratio-posteriors}. In total, we constrain six lensing ratios at the 13-23\% level. The highest signal-to-noise ratio constraints are those corresponding to the highest lens redshift bin.  

We first fit the measured ratios using a fiducial cosmological model.  We compute the expectation value of the ratios using the best-fit cosmology from the \texttt{TT,TE,EE+lowP+lensing+ext} analysis in \citet{PlanckCollaboration2015}. We call these values $r_{\rm Planck}$ and fit the measured ratios with a model of the form $\hat{r} = A r_{\rm Planck}$, where $A$ is a free parameter.  We find $A = 1.1 \pm 0.1$.  This measurement demonstrates that the ratio measurements are consistent with the fiducial cosmology within the statistical error bars, and are measured a combined precision of roughly 10\%.  For comparison, the measurement of lensing ratios presented in \citet{Miyatake:2017} using CMASS galaxies as tracers, galaxy shapes from CFHTLenS and \textit{Planck} data, reports a 17\% uncertainty on a joint measurement of the ratio, obtained from combining results from three tracer galaxy redshift bins and a single source galaxy bin.  The $\chi^2$ per degree of freedom for the measurements relative to the $r_{\rm Planck}$ model is $8.5/5$, corresponding to a probability to exceed of 0.13.  This indicates a reasonable fit to the {\it Planck} model.  Note that the ratio measurements for different source bins but the same tracer galaxy bin are highly covariant, as can be seen in Fig.~\ref{fig:ratio-posteriors}.

\subsection{Cosmological constraints}

We now use the ratio measurements presented above to constrain cosmological parameters.  As an illustration of the cosmological sensitivity of the ratios, Fig.~\ref{fig:miyatakestyle} shows the theoretical predictions for two cosmological $\Lambda$CDM models with different values of $\Omega_k$, with $\Omega_{\rm m}$ fixed to 0.3.  Throughout this analysis, we fix the redshift of the surface of last scattering to $z^* = 1090$.  From this figure, we see that negative values of $\Omega_k$ have a significantly greater impact on the lensing ratios than positive values. This is due to the fact that the angular diameter distance to the surface of the last scattering changes more with curvature for negative $\Omega_k$ than for positive $\Omega_k$.  

To obtain cosmological constraints we use the methodology described in Sec.~\ref{sec:complete_model}. We consider curved $\Lambda$CDM models where we vary the cosmological parameters $\Omega_{ k}$ and $\Omega_{\rm{m}}$ and the systematics parameters described in Sec.~\ref{sec:sys_model}. We use the priors on the multiplicative shear bias derived in \citet{shearcat} and the redshift bias parameters from \citet{Hoyle2018,Davis2017,Gatti2018}. 

\begin{figure}
\centering
\includegraphics[width=0.47\textwidth]{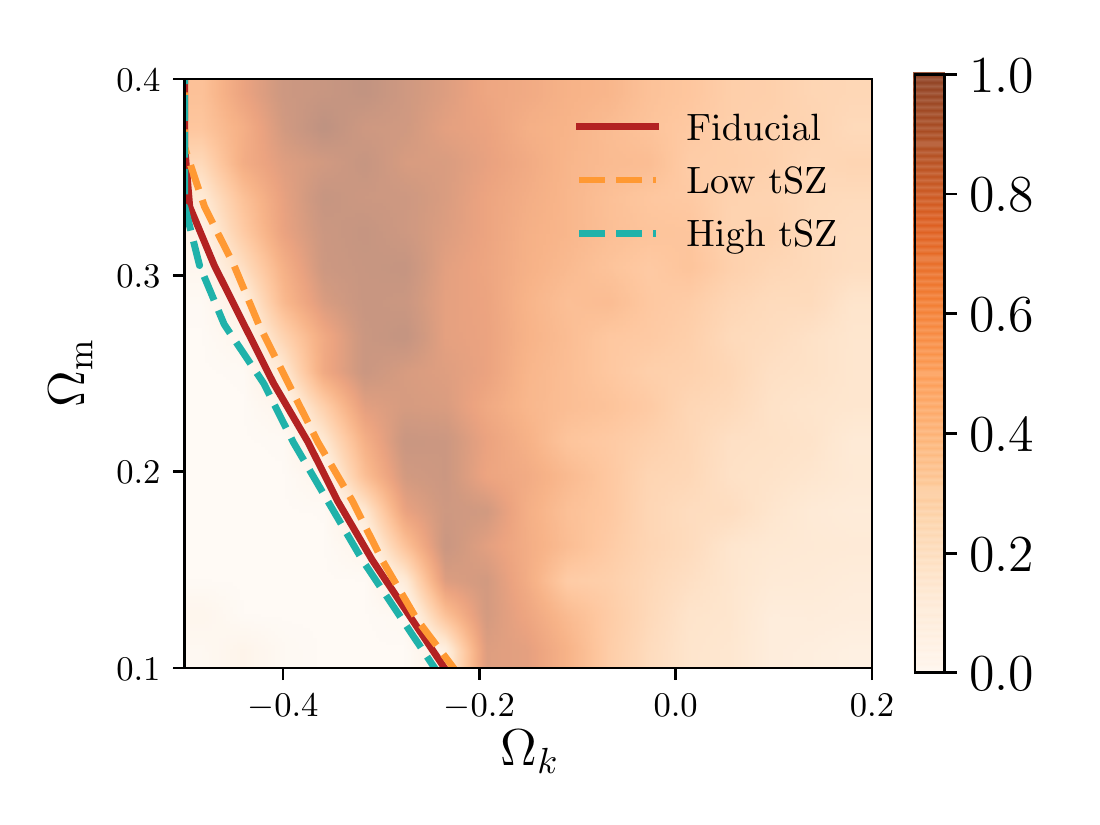}
    \caption{The constraints on $\Omega_{\rm m}$ and $\Omega_k$ resulting from analysis of the measured lensing ratios.  The background color shows the marginalized posterior density for these two parameters.  The data strongly rule out regions of parameter space with low $\Omega_{\rm m}$ and very negative $\Omega_k$.  At each $\Omega_{\rm m}$, we identify a lower limit on $\Omega_k$ by identifying the value of $\Omega_k$ for which the marginalized posterior falls by $1/e^2$ relative to the maximum, which for a Gaussian distribution would correspond to the $2\sigma$ lower limit.  This limit is illustrated with the red solid curve. We also show (dashed curves) the changes to these limits when using two variations on the fiducial tSZ model, as described in Sec.~\ref{sec:tsz_correction}. }
    \label{fig:data_constraints}
\end{figure}

Fig.~\ref{fig:data_constraints} shows the resultant marginalized posterior density (colored region) as a function $\Omega_{\rm m}$ and $\Omega_k$.  We find that the data strongly rule out low values of $\Omega_{\rm m}$ and very negative values of $\Omega_k$.  However, at each $\Omega_{\rm m}$, we obtain only a lower limit on $\Omega_k$.  Consequently, we focus on how the data constrain $\Omega_{k}$.   We derive limits on $\Omega_k$ in the following way.  For each value of $\Omega_{\rm m}$, we determine the value of $\Omega_k$ such that the marginalized posterior on $\Omega_k$ is lower than the peak of the posterior by a factor of $1/e^2$.  For a Gaussian distribution, this would correspond to the $2\sigma$ lower limit. This limit is shown in Fig.~\ref{fig:data_constraints} as the solid red line.  Consistent with the marginalized posterior, we rule out very negative $\Omega_k$, with the limit tightening for lower values of $\Omega_{\rm m}$. 

As seen in Fig.~\ref{fig:data_constraints}, the data somewhat prefer models with negative curvature over models with $\Omega_k = 0$.  This preference is driven by the high redshift data points seen in Fig.~\ref{fig:miyatakestyle}.  However, this preference is not statistically significant.  For $\Omega_k \gtrsim -0.1$, the posterior on $\Omega_k$ is quite flat for all $\Omega_m$.  This is consistent with the finding noted above that the amplitude of the lensing ratios is consistent (to $1\sigma$) with the prediction from flat $\Lambda$CDM, which has $\Omega_k = 0$.

Fig.~\ref{fig:data_constraints} also shows the impact of using the high and low-amplitude tSZ models  (see discussion in Sec.~\ref{subsec:tsz_validation}) on the cosmological constraints with the green and orange dashed curves, respectively.  The uncertainty on the tSZ amplitude contributes a non-negligible amount of systematic uncertainty to our analysis, but it is subdominant to the statistical uncertainty. 

We have tested that the constraints obtained by varying only the cosmological parameters, and not marginalizing over the shear and photometric redshifts systematics parameters are essentially identical to those obtained when the systematics parameters are varied. Therefore, we conclude that at the current level of statistical uncertainty on the lensing ratios, the impact of systematics errors in photometric redshifts and shear calibration are not significant.  Note that the systematics parameters $\Delta z_s^i$, $\Delta z_l^i$, and $m_i$ are strongly prior dominated. 

\section{Forecasts}
\label{sec:projections}

Upcoming data from DES, SPT and future surveys have the potential to significantly reduce the statistical uncertainty on measurements of lensing ratios. \citet{Das:2009} calculated the uncertainty on lensing ratios that could be obtained with the combination of a lens galaxy sample from a futuristic spectroscopic survey, a LSST-like galaxy weak lensing survey, and a CMB lensing map from a CMBPOL-like survey.  They found that a roughly 1\% constraint on the lensing ratio could be obtained with this combination of experiments, and that such a constraint could contribute useful cosmological information that is complementary to e.g. \textit{Planck} and future measurements of the baryon acoustic oscillation (BAO) feature in the galaxy distribution.  

Here, we extend the analysis of \citet{Das:2009} to account for the effects of systematic errors in the redshift and shear measurements.  We also update the forecasts given current expectations for future survey designs.  Finally, we show how using a lens galaxy population identified with photometric data from LSST can be used to decrease the error bars on the ratio measurements.  For this analysis, we consider curved $w$CDM cosmological models, parameterized by $\Omega_{\rm m}$, $\Omega_k$ and $w$, the equation of state parameter of dark energy.

As discussed in Sec.~\ref{sec:model}, there are several potential sources of systematic error that could affect measurement of lensing ratios beyond errors in the source redshift distributions and shear calibration errors.  In particular, tSZ bias in the $\kappa_{\rm CMB}$ maps is a potentially significant concern.  Here, we ignore bias due to tSZ contamination of the $\kappa_{\rm CMB}$ map under the assumption that future experiments will use lensing estimators based on CMB polarization data (which is much less severely impacted by tSZ), or that they will use some multi-frequency cleaning strategy, such as that discussed in \citet{Madhavacheril:2018}.

\subsection{Calculation of projected uncertainty}
\label{sec:uncertainty_calculation}

\begin{figure}
\centering	\includegraphics[width=0.45\textwidth]{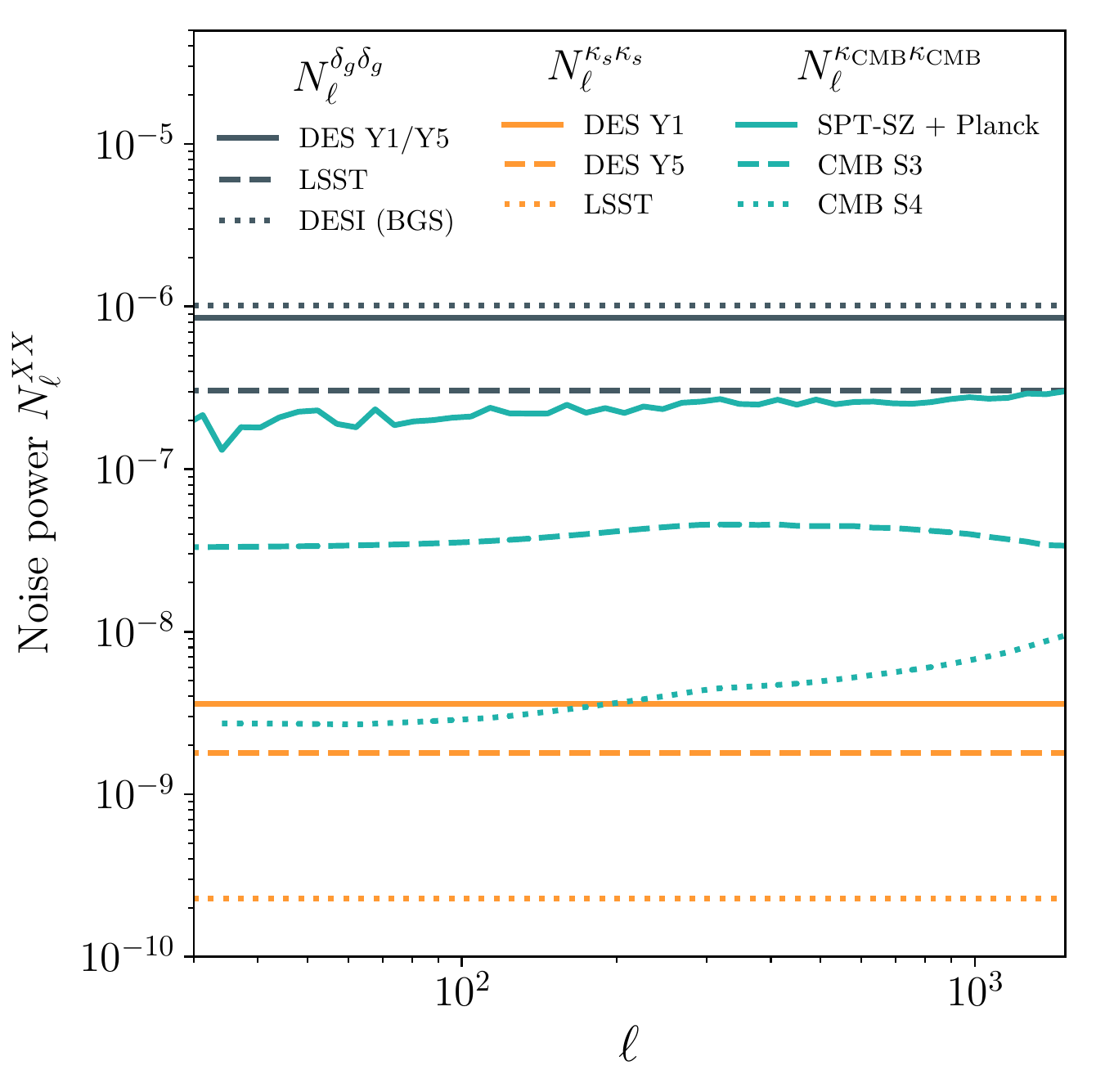}
    \caption{Noise power spectra for the experimental configurations described in Sec.~\ref{sec:experiment_config}.  The noise power for $C_{\ell}^{\delta_g\delta_g}$ (gray curves) is inversely proportional to the lens galaxy density; the noise power for $C_{\ell}^{\kappa_{s} \kappa_{s}}$  (orange curves) is inversely proportional to source galaxy denisty.  Noise power in the CMB lensing maps (blue curves) is dependent on the details of the experimental configurations of the CMB telescope.  For SPT-SZ we use the measured CMB lensing noise power, while for CMB-S3 and S4, we use forecasts.}
    \label{fig:noise_power}
 \end{figure}

To estimate the error on the lensing ratios with future data we use a methodology similar to \citet{Das:2009}.  We define 
\begin{equation}
\label{eq:zl}
Z_{\ell} = C_{\ell}^{\kappa_{\rm CMB}\delta_g} - r\, C_{\ell}^{\kappa_{s}\delta_g},
\end{equation}
and a corresponding $\chi^2$ via
\begin{equation}
\label{eq:chi2}
\chi^2(r) = \sum_l \frac{Z_l^2}{\sigma^2(Z_l)},
\end{equation}
where $\sigma^2(Z_l)$ is the variance of $Z_l$.   The uncertainty on the ratio, $\sigma(r)$, can then be calculated as
\begin{equation}
\label{eq:sigmar}
\frac{1}{\sigma^2(r)} = \frac{1}{2} \frac{\partial^2 \chi^2(r)}{\partial r^2}.
\end{equation}
To compute $Z_l$, we must extend the formalism of \citet{Das:2009} to include partial overlap between surveys.  Given a fiducial value of the ratio, $r_0$, the variance of $Z_l$ can be computed using the expressions in \citet{White:2009}. We find 
\begin{equation}
\label{eq:sigmaz}
\begin{split}
\sigma^2\left(Z_{\ell}\right) =& \frac{1}{(2\ell+1)} \left[ \frac{1}{f_{\rm sky}^{\kappa_{\rm CMB}\delta_g}} \left( \tilde{C}_\ell^{\kappa_{\rm CMB}\kappa_{\rm CMB}} \tilde{C}_\ell^{\delta_g\delta_g} + \left(C_{\ell}^{\kappa_{\rm CMB}\delta_g}\right)^2 \right) \right. \\
&+ \frac{r_0^2}{f_{\rm sky}^{\kappa_{s}\delta_g}} \left( \tilde{C}_\ell^{\kappa_{s}\kappa_{s}} \tilde{C}_\ell^{\delta_g\delta_g} + \left(C_{\ell}^{\kappa_{s}\delta_g}\right)^2  \right) \\
&-2r_0\frac{f_{\rm sky}^{\kappa_{\rm CMB} \kappa_{s} \delta_g} }{f_{\rm sky}^{\kappa_{\rm CMB} \delta_g} f_{\rm sky}^{\kappa_{s}\delta_g}} \left.\left( C_\ell^{\kappa_{\rm CMB}\kappa_{s}} \tilde{C}_\ell^{\delta_g\delta_g} +  C_\ell^{\kappa_{\rm CMB}\delta_g} C_\ell^{\kappa_{s}\delta_g} \right)\, \right],
\end{split}
\end{equation}
where
\begin{equation}
\tilde{C}_\ell^{XX} = C_\ell^{XX} + N_\ell^{XX},
\end{equation}
and $N_\ell$ is the corresponding noise power spectrum.  The Poisson noise for the tracer sample is $N_\ell^{\delta_g\delta_g} = 1/n_g$, where $n_g$ is the number density of tracer galaxies per steradian.  We compute $N_\ell^{\kappa_{s}\kappa_{s}}$ as 
\begin{equation}
N_\ell^{\kappa_{s}\kappa_{s}}= \frac{\sigma_\epsilon^2}{n_{\rm{s}}},
\end{equation}
where $\sigma_\epsilon$ is the standard deviation of the weighted galaxy shapes and $n_{\rm{s}}$ is the number density of the source galaxies per steradian used to produce the lensing maps.  We adopt $\sigma_{\epsilon} = 0.26$ below.   The various noise curves used in the forecasts are shown for the different surveys in Fig.~\ref{fig:noise_power}.

The $f_{\rm sky}$ factors in Eq.~(\ref{eq:sigmaz}) approximately take into account the fact that the variance of the $C_{\ell}$ measurements is increased for partial sky coverage.  We define $f_{\rm sky}^{\kappa_{\rm CMB} \delta_g}$ and $f_{\rm sky}^{\kappa_{s} \delta_g}$ as the sky fractions over which $C_{\ell}^{\kappa_{\rm CMB}\delta_g}$ and $C_{\ell}^{\kappa_{s}\delta_g}$ are measured, respectively, and $f_{\rm sky}^{\kappa_{\rm CMB} \kappa_{s} \delta_g}$ is the sky fraction over which the $\delta_g$, $\kappa_{\rm CMB}$ and $\kappa_{s}$ measurements all overlap. In the case that there is no overlap between all three measurements, $C_{\ell}^{\kappa_{\rm CMB}\delta_g}$ and $C_{\ell}^{\kappa_{s}\delta_g}$ are uncorrelated and the variance of $Z_l$ is given by the sum of the variances of the two terms in Eq.~(\ref{eq:zl}).  In the case where there is overlap between the lens galaxies, source galaxies, and CMB lensing measurements, some reduction of variance can be obtained via sample variance cancellation.

Finally, $\sigma(r)$ is calculated by substituting Eqs.~(\ref{eq:chi2}) and (\ref{eq:sigmaz}) into Eq.~(\ref{eq:sigmar}). For the purposes of these forecasts, we adopt the best-fit $\Lambda$CDM cosmological model from the analysis of \texttt{TT,TE,EE+lowP+lensing+ext} datasets in \citet{PlanckCollaboration2015}.

\begin{table*}
\begin{center}
\begin{tabular}{ l c c c c c c }
\toprule
  Surveys & Lens $z$ range & $N_{\mathrm{lens\ bins}}$ & Source $z$ range  & $N_{\mathrm{source \ bins}}$ & $\sigma_{r, \rm stat}$ [min, max]  \\
   \midrule 
    DES Y1 + SPT-SZ (current measurements) & $0.15<z_l<0.6$ & 3 &  $0.6<z_s<1.3$ & 2 &[0.13, 0.23]\\ 
   DES Y5 + SPT-SZ & $0.15<z_l<0.6$ & 3 &  $0.6<z_s<1.3$ & 2 &[0.098, 0.15]\\ 
 DES Y5 + CMB-S3 & $0.15<z_l<0.6$ & 3 &  $0.6<z_s<1.3$ & 2 &[0.042, 0.060]\\ 
DESI + LSST + CMB-S4 &  
\begin{tabular}{@{}c@{}}$0.2<z_l<0.4$ (BGS)  \\  $0.8 < z_l < 1.0$ (ELG) \end{tabular}
& 
\begin{tabular}{@{}c@{}}4 \\ 2\end{tabular}
& $1.0 < z_s <1.6$ & 1 & 
\begin{tabular}{@{}c@{}} $[0.018, 0.019]$ (BGS) \\ $[0.040, 0.054]$ (ELG)\end{tabular}
\\
LSST + CMB-S4 & $0.2 < z_l < 0.7$ & 10 & $1.0 < z_s <1.6$ & 1 & [0.013,0.015] \\
\bottomrule
\end{tabular}
\end{center}
\caption{Forecasts for precision of ratio measurements for the future experiment configurations described in Sec.~\ref{sec:experiment_config}, except for the first row, which corresponds to the measurements presented in this paper in Fig.~\ref{fig:miyatakestyle}.  Last column represents the minimum and maximum statistical errors on the ratios over all tracer and source galaxy bin combinations.  \label{table:error_surveys}}
\end{table*}

\subsection{Future experiment configuration}
\label{sec:experiment_config}

We consider several future experimental configurations using both current and future surveys, which are also summarized in Table~\ref{table:error_surveys}:
\begin{itemize}
\item {\bf DES Y5 + SPT-SZ}: this represents what can be achieved with full-survey DES data and current SPT-SZ data.  We assume an overlapping area of 2500 sq. deg. (i.e. the full area of the SPT-SZ survey).  For the tracer and source galaxies, we adopt the current redshift bins and the same number densities for the tracer galaxies; we assume an increased source density of a factor of two with respect to the Y1 density, due to the higher depth of Y5 data. The assumed CMB noise power, $N_\ell^{\kappa_{\rm CMB}\kappa_{\rm CMB}}$, is taken from \citet{Omori2017}.  Finally we assume that tSZ bias can be mitigated using multi-frequency information, allowing us to exploit all angular scales. 

\item {\bf DES Y5 + Stage 3 CMB}: this represents what can be achieved with full-survey DES data and a near-term, Stage 3 CMB experiment (CMB-S3).  Stage 3 CMB experiments include SPT-3G \citep{Benson:2014} and Advanced ACTPol \citep{Henderson:2016}.  We assume an overlapping area of 5000 sq. deg. and use the CMB-S3 noise curve from \citet{CMBS4}.  We adopt the same tracer and source galaxy bins as the current analysis, with a source density of twice the Y1 density.  

\item {\bf DESI + LSST + Stage 4 CMB}: this represents one possible use of future survey data to constrain lensing ratios.  We assume that the tracer galaxies are spectroscopically identified using the Dark Energy Spectroscopic Instrument \citep{DESI}, allowing us to ignore redshift errors for this sample.  The tracer galaxies are assumed to be drawn from two DESI populations: a set of low-$z$ galaxies from the Bright Galaxy (BGS) sample and a set of high-$z$ galaxies from the Emission Line Galaxy (ELG) sample.  The BGS tracer galaxies are divided into four redshift bins between $z=0.2$ and $z=0.4$, and the tracer galaxy bias is assumed to be $1.34/D(z)$, where $D(z)$ is the linear growth factor, normalized to $D(z=0)=1$; the ELG galaxies are divided into two redshift bins between $z=0.8$ and $z=1.0$, and are assumed to have a bias of $0.84/D(z)$ \citep{DESI}. The tracer galaxy density for the BGS redshift bins (width of $\Delta z = 0.05$) is assumed to be 75 per sq. deg. and 150 per sq. deg for the ELG redshift bins (width of $\Delta z = 0.1$).  We assume that LSST \citep{LSST} is used to measure shapes of source galaxies, with a source density of 25 galaxies per sq. arcmin and redshift range from $z=1.0$ to $z=1.6$. 

The CMB lensing map is assumed to come from a Stage 4 (CMB-S4) like experiment \citep{CMBS4}; we adopt the minimum variance CMB lensing noise curve from \citet{Schaan:2017}.  Finally, we assume overlapping area between DESI and CMB-S4 of 16500 sq. deg., overlap between DESI and LSST of 3000 sq. deg., and overlap between all three surveys of 3000 sq. deg. 

\item {\bf LSST + CMB-S4}: another possible use of future survey data for measuring lensing ratios is to define a tracer galaxy sample using photometric data from LSST.  As we have shown above, algorithms like redMaGiC can be used to define galaxy populations that are sufficiently narrowly distributed in redshift for the purposes of measuring lensing ratios.   We assume that the LSST tracer galaxy sample is divided into 10 bins between $z=0.2$ and $z=0.7$, with number density of 100 galaxies per square degree for each bin.  Such densities are comparable to what is currently achieved with DES redMaGiC.  We make the same source galaxy and CMB lensing assumptions as above.
\end{itemize}

In addition to the survey assumptions described above, we must adopt some prescription for the expected systematic errors on shear calibration and photometric redshift determination.  We assume that the multiplicative shear bias from LSST can be calibrated to $\sigma(m) = 0.001$, which is the requirement set in \citet{LSST:ScienceBook} and also of the order of what is expected from \citet{Schaan:2017}.  When using DESI to create the tracer galaxy sample, we ignore redshift errors in the analysis; for LSST we assume that with a redMaGic-like algorithm, the tracer galaxy redshifts can be calibrated to $\sigma(\Delta z_l) = 0.005$.  We assume that the source photo-$z$s measured by LSST can be calibrated to the level of $\sigma(\Delta z_s) = 0.01$ \citep{LSST:ScienceBook}.

Note that in the forecasts below, we ignore the issue of the finite width of the tracer galaxy redshift bins.  For the survey assumptions defined above, we have tested that the errors on the ratios induced by the narrow lens approximation are significantly below the statistical uncertainties on the ratios.  Furthermore, given the small assumed redshift errors of the lens galaxies, we could in principle divide the tracer galaxies into more redshift bins and the narrow lens approximation would improve.  We find, however, that doing so does not appreciably change our results.

\begin{figure}
\centering	\includegraphics[width=0.5\textwidth]{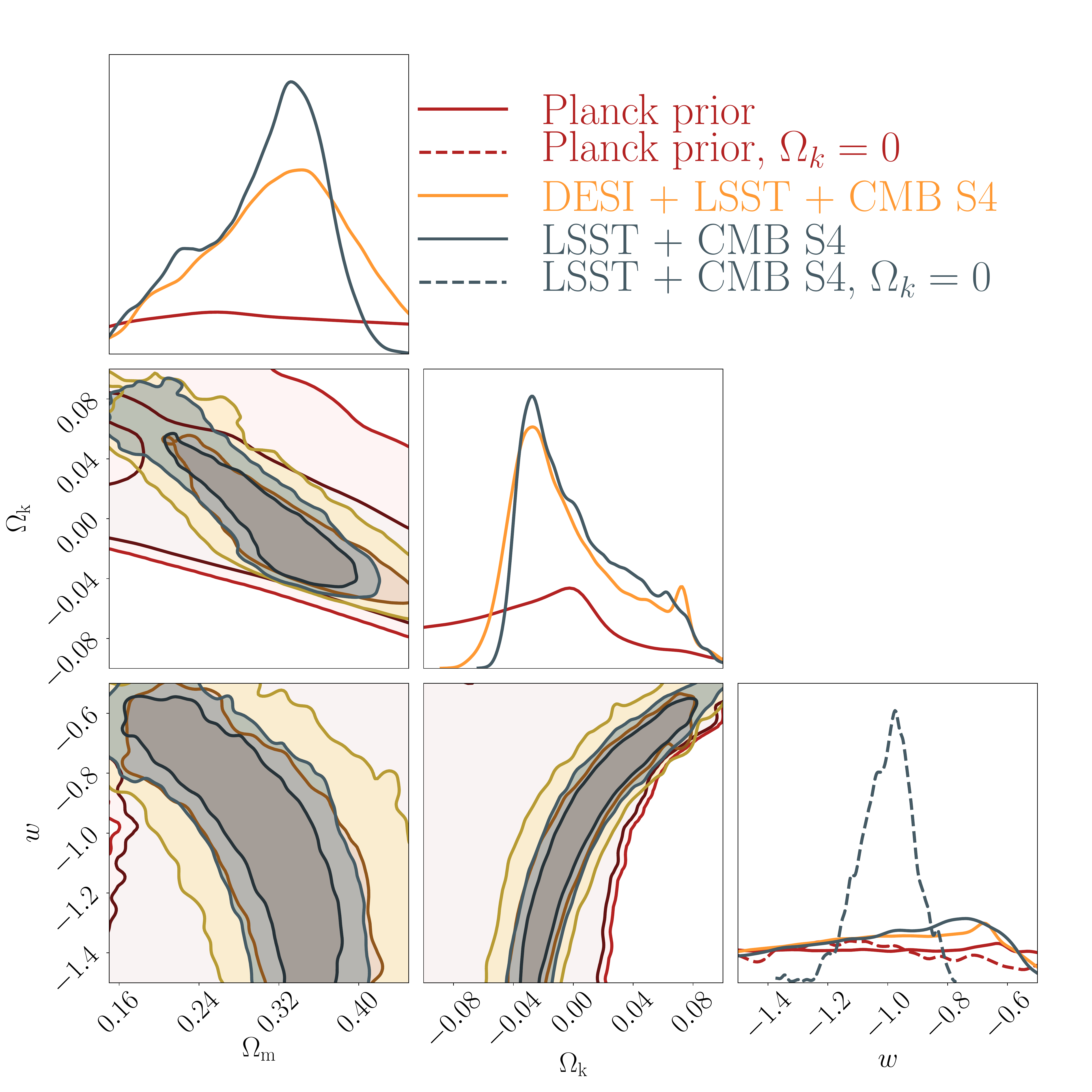}
    \caption{Projected cosmological constraints from lensing ratios when using LSST + DESI + CMB-S4 vs. LSST + CMB-S4, using the geometrical \textit{Planck} prior, which is also shown in the figure.  We have marginalized over parameters describing systematic uncertainties in lens and source galaxy redshifts, and systematic errors in source galaxy shears.  We also marginalize over $h$ and $\Omega_b$ as these appear in the geometrical {\it Planck} prior (see text).  The constraints that can be obtained using a photometrically identified tracer galaxy population (LSST+CMB-S4) are tighter than those that can be obtained from a spectroscopically identified tracer galaxy population (DESI+LSST+CMB-S4).  Apparently, the increased number density of the tracer galaxies with the photometric survey outweighs the increased redshift uncertainties.}
    \label{fig:future_compare}
\end{figure}

\subsection{Future constraints on lensing ratios}

There are three sources of statistical noise in the measurements of the lensing ratios: noise in the measurement of galaxy density, noise in the galaxy lensing maps, and noise in the CMB lensing maps. For current data, all of these components make significant contributions to the total uncertainty on the ratios, although noise in the CMB lensing map and galaxy density dominate.  For instance, increasing the number density of tracers by a factor of two would decrease the uncertainty on the ratios by roughly $15\%$. Significant improvement could also be obtained by reducing the noise in the $\kappa$ maps, especially $\kappa_{\rm CMB}$. Halving the noise in the CMB $\kappa$ maps would decrease the ratio uncertainty by $25\%$, while the same improvement in the galaxy $\kappa$ maps would reduce the ratio uncertainty by $5\%$. Finally, doubling the area of the surveys would reduce by $40\%$ the uncertainty on the ratios. The future survey configurations described in Sec.~\ref{sec:experiment_config} make improvements to the lensing ratio constraints in all of these ways.

The projected cosmological constraints on $\Omega_{\rm m}$, $\Omega_{ k}$ and $w$ obtained from the forecasted lensing ratio constraints for DESI, LSST and CMB S4 are shown in Fig.~\ref{fig:future_compare}, assuming the tomographic ratio measurements are independent. This is a reasonable assumption because for these configurations there is only one source bin (see Table.~\ref{table:error_surveys}) and the covariance between measurements using different tracer bins is small, as shown in Fig.~\ref{fig:ratio-posteriors}. When generating this figure, we have adopted priors from the \textit{Planck} measurement of the CMB power spectrum in \citet{PlanckCollaboration2015}. Since the lensing ratio measurements are purely geometrical in nature, we choose to use only geometric information from the CMB power spectrum.  For this purpose, we use the geometric CMB prior defined in \citet{BOSS:2015}.  Since most of the information in this prior comes from the first few peaks of the CMB temperature power spectrum, constructing this prior from the \textit{Planck} constraints is a reasonable approximation for future surveys. Since the CMB prior depends on $h$ and $\Omega_{\rm b}$, we have marginalized over these quantities in generating Fig.~\ref{fig:future_compare}.  Additionally, in Fig.~\ref{fig:future_compare} we have marginalized over the systematics parameters $\Delta z_l$, $\Delta z_s$ and $m$ for each redshift bin, imposing the priors described in Sec.~\ref{sec:experiment_config}.

Fig.~\ref{fig:future_compare} makes it clear that the lensing ratios contribute information beyond that contained in the geometrical CMB prior.  Because of the "geometrical degeneracy" in the CMB power spectrum \citep{Efstathiou:1999}, the CMB constraints on $\Omega_{\rm m}$, $\Omega_{ k}$ and $w$ are quite weak when all three parameters are varied simultaneously (the red contours).  However, future lensing ratio constraints help to break these degeneracies.  The combination of the lensing ratio and geometric CMB prior is particularly powerful in the space of $\Omega_k$ and $w$.  The main impact of the lensing ratio constraints is to remove regions of parameter space with negative $\Omega_k$ and with small $w$ in absolute value, leading to a tight degeneracy between $\Omega_k$ and $w$.  This degeneracy can be broken using e.g. information from BAO \citep{Das:2009}.  Alternatively, if flatness is assumed (i.e. $\Omega_{ k} = 0$), the resultant constraint is $w = -1.0 \pm 0.1$ (grey dashed curve in lower right panel). 

Additionally, from Fig.~\ref{fig:future_compare} it can be seen that the cosmological constraints obtained from using LSST redMaGiC-like galaxies as the tracers are tighter than what is obtained by using DESI galaxies as the tracers.  This is one of the main findings of our analysis: because of the tight photometric redshift errors that can be obtained with a redMaGiC-like algorithm, lensing ratios can be measured to high precision using a combination of photometric galaxy measurement and CMB lensing.  A spectroscopic lens galaxy catalog is not necessary for the purposes of measuring lensing ratios.  For fixed $w = -1$, the constraint obtained on $\Omega_k$ for the case of DESI tracers is $\sigma(\Omega_k) = 0.014$; for the case of LSST tracers, it is $\sigma(\Omega_k) = 0.009$.  Similarly, for fixed $\Omega_k = 0$, the constraint obtained on $w$ for the case of DESI tracers is $\sigma(w) = 0.15$; for the case of LSST tracers, it is $\sigma(w) = 0.09$.

\begin{figure}
\centering
		\includegraphics[width=0.4\textwidth]{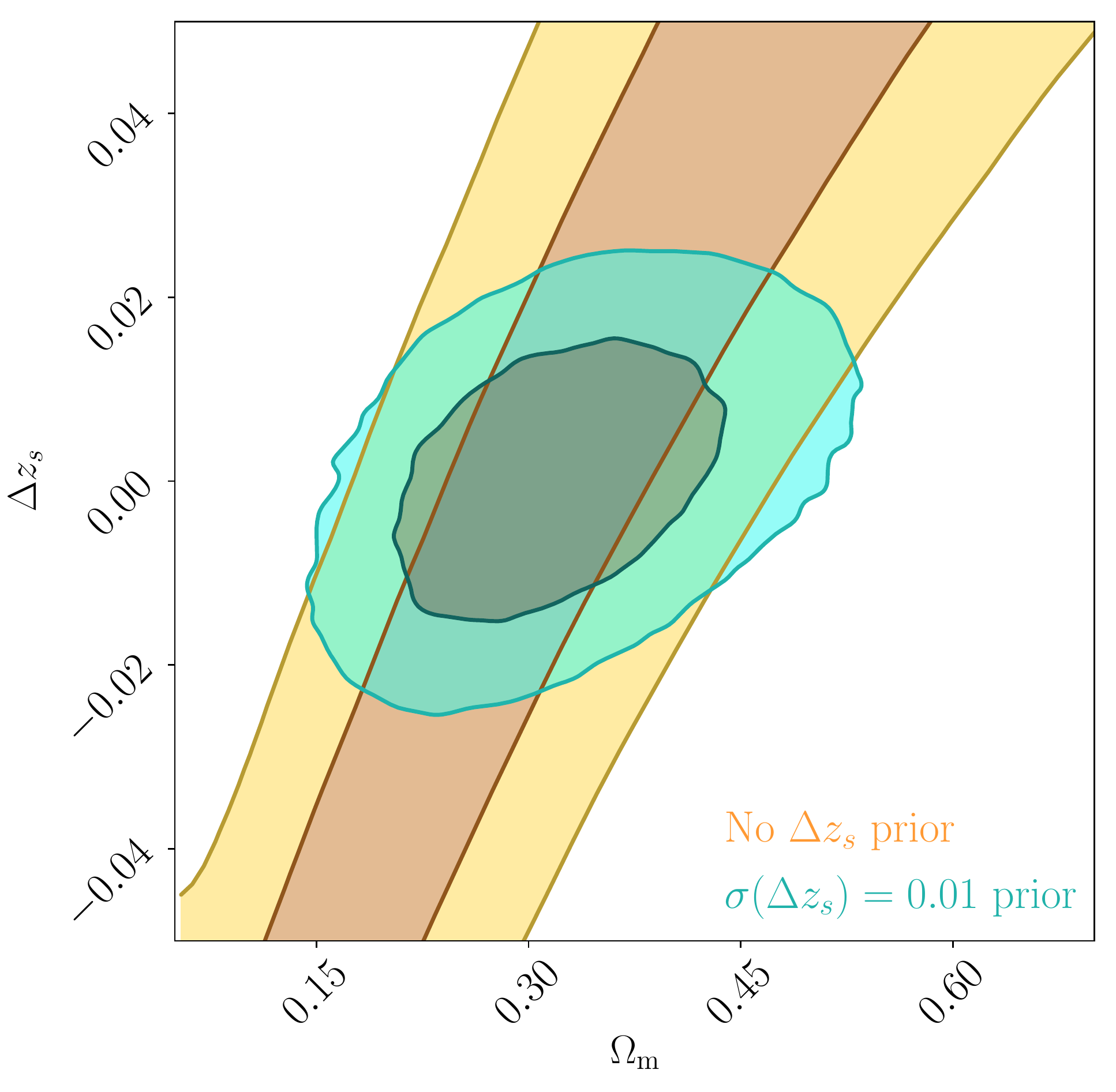}
    \caption{Degeneracy between $\Omega_{m}$ and source redshift bias, $\Delta z_s$, for the case of one lens and source redshift bin when the lensing ratio is measured to 1\% precision.  Since there is only one ratio measurement, the constraint on $\Omega_{\rm m}$ is completely degenerate with the redshift bias.}
    \label{fig:om_sys}
\end{figure}

\begin{figure}
\centering
\includegraphics[width=0.4
\textwidth]{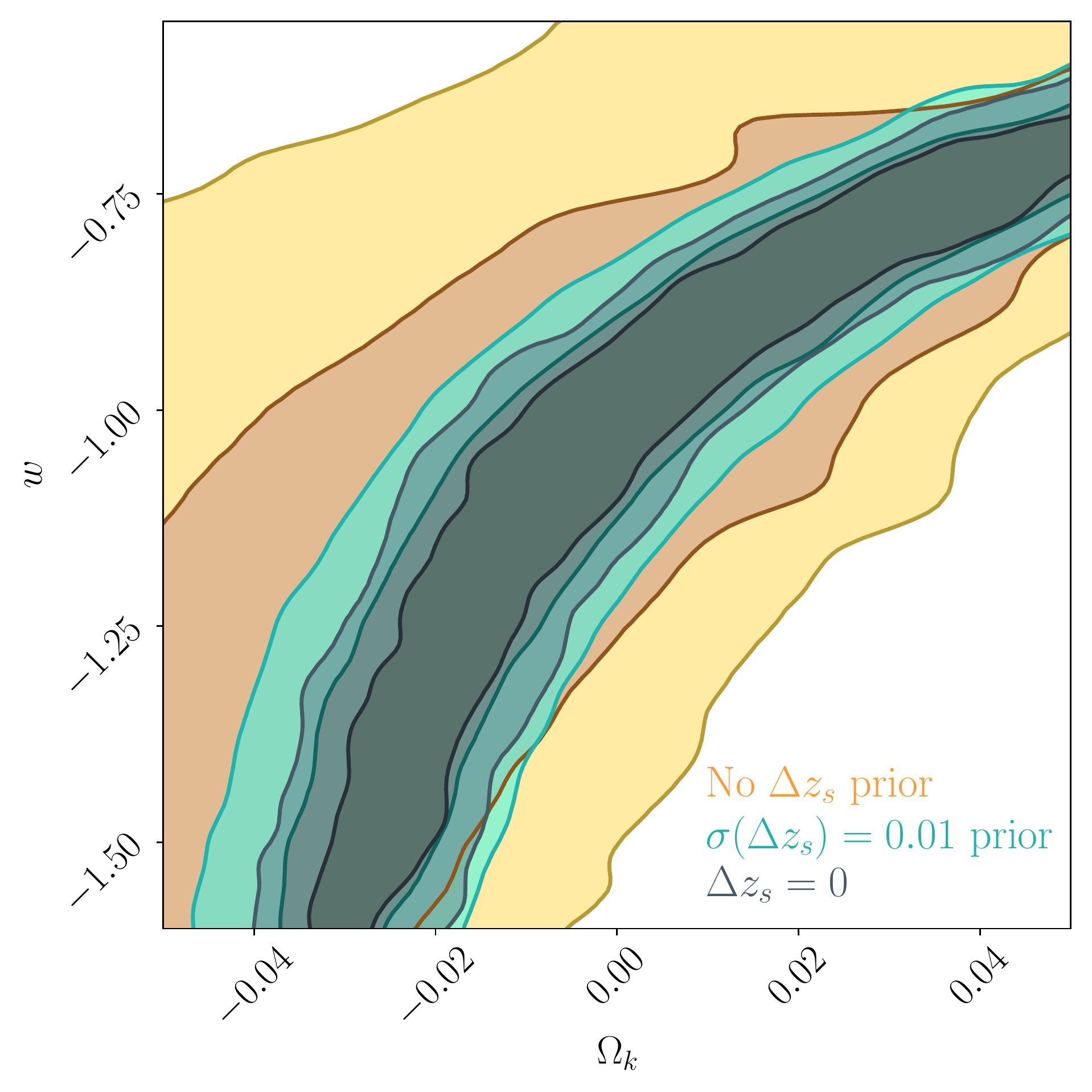}
    \caption{Projected constraints on $\Omega_k$ and $w$ for different priors on the redshift bias parameter, $\Delta z_s$.  We have assumed the projected constraints for LSST+CMB S4 in this figure.  Uncertainty on $\Delta z_s$ significantly degrades the cosmological constraints.  For the projected level of constraints, $\sigma(\Delta z_s) = 0.01$, the degradation is small, but non-zero.
    }
    \label{fig:bigfig}
\end{figure}

\begin{figure}
\centering
\includegraphics[width=0.5\textwidth]{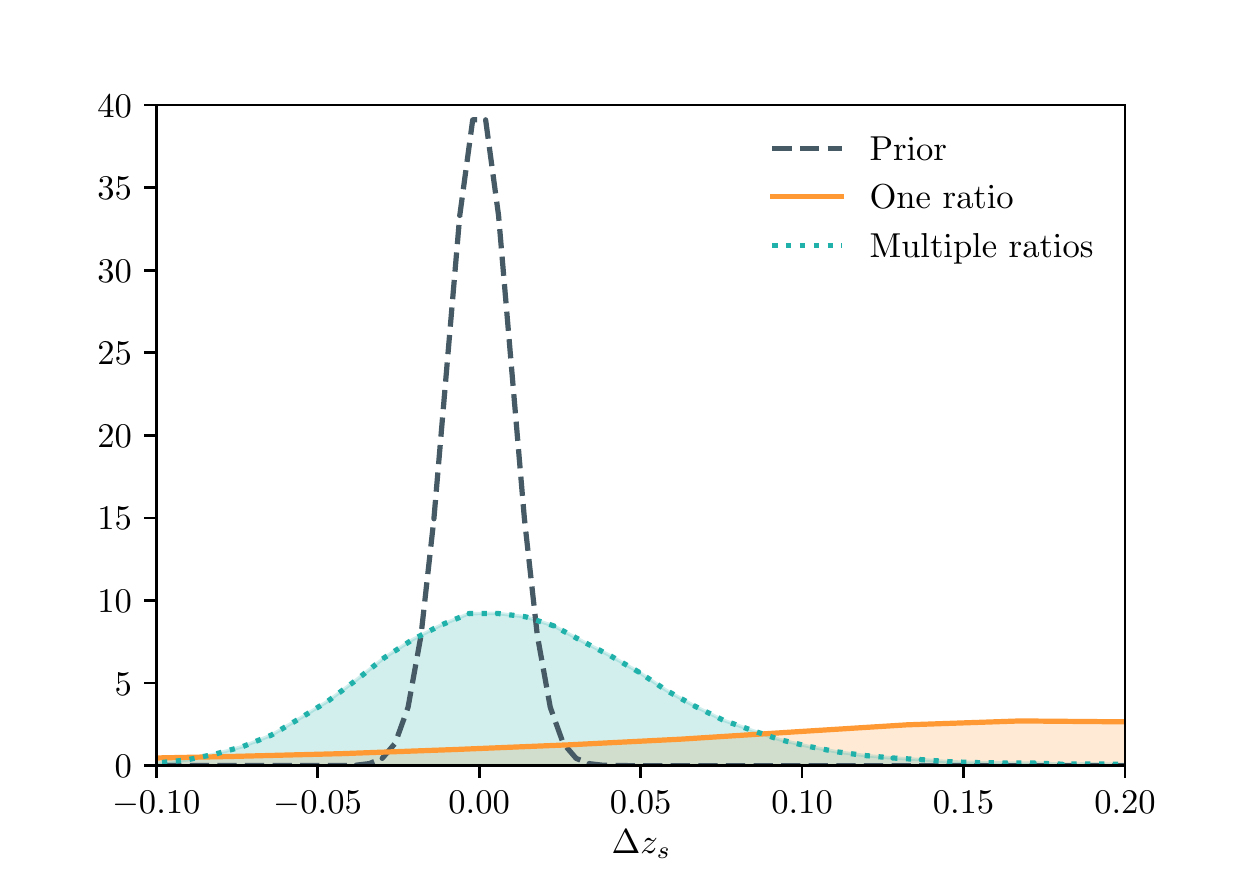}
    \caption{Posteriors on the source redshift bias parameter, $\Delta z_s$, when using only one lens bin and source bin (one ratio) vs using three lens and one source bins (three ratios).  Using multiple lens redshift bins allows one to obtain some self-calibration of the photo-$z$ bias.  However, the level of self-calibration achieved is not as tight as expected priors, $\sigma(\Delta z_s) = 0.01$. }
    \label{fig:sys_calibration}
\end{figure}

\subsection{Impact of systematic errors on lensing ratios}\label{sec:projections-sys-errors} 

We now investigate in more detail the impact of systematic errors on future ratio measurements.  For illustrative purposes, we first consider the case of a ratio measurement using a single lens and source galaxy bin, for which we adopt a 1\% error typical of the LSST + CMB-S4 forecasts.  In this case, since there is only a single ratio measurement, systematic errors on shear calibration and photometric redshift bias will be completely degenerate with the cosmological constraints.
 This degeneracy is illustrated for the case of $\Omega_{\rm m}$ in Fig.~\ref{fig:om_sys}.  Without a prior on $\Delta z_s$,  $\Omega_{\rm m}$ cannot be constrained at all (orange contour).  Given the projected prior on $\Delta z_s$ of 0.01, we can obtain a constraint on $\Omega_{\rm m}$ (light green contour).  However, in this case the cosmological constraint will be strongly determined by the accuracy of our prior on $\Delta z_s$ and the constraining power on  $\Omega_{\rm m}$ will be reduced by photo-$z$ uncertainties. 
 
In Fig.~\ref{fig:bigfig} we show the impact of source redshift errors in the $w-\Omega_k$ plane when more cosmological parameters are added into the model, and also when multiple ratios coming from different lens and source bin combinations are used. For this figure, lensing ratio constraints are taken from LSST + CMB-S4 forecasts, which match the ones from Fig.~\ref{fig:future_compare} for the contours marginalizing over $\Delta z_s$ with the fiducial prior (light blue). In this figure, we also show the results of marginalizing over $\Delta z_s$ with a wide, flat prior (orange), and fixing $\Delta z_s = 0$ (grey). Degeneracy between the cosmological parameters and $\Delta z_s$ can have a large impact, as evidenced by the change in constraints in going from $\Delta z_s = 0$ to marginalizing over $\Delta z_s$ with the wide, flat prior.  However, with the assumed $\Delta z_s$ priors of 0.01 we find that the effect of source redshift uncertainty on the lensing ratios constraints is fairly small, but also not negligible. 

If multiple lens redshift bins are used to measure multiple lensing ratios, the degeneracy between $\Delta z_s$ and the cosmological parameters can be broken somewhat. To illustrate this point, Fig.~\ref{fig:sys_calibration} shows the posteriors on $\Delta z_s$ when using a single ratio or multiple ratio measurements.   With only a single ratio measurement (orange curve), the ratio is highly degenerate with the systematic uncertainty on the redshift bias parameter, $\Delta z_s$, so the posterior on $\Delta z_s$ is very broad.  Using multiple ratios allows for some self-calibration of the photo-$z$ bias (blue curve); in this case, the ratio measurements alone are being used to calibrate $\Delta z_s$. However, we see that the level of self-calibration of $\Delta z_s$ remains weaker than the prior (black dashed curve) and, therefore, not using any $\Delta z_s$ prior in the cosmology analysis would result in some degradation of the cosmology constraints. Note that the preference for large $\Delta z_s$ exhibited in Fig.~\ref{fig:sys_calibration} for the case of a single ratio measurement is due to the projection of the higher-dimensional  parameter space to the one-dimensional constraints on $\Delta z_s$.

We have also investigated the impact of shear calibration uncertainty on the constraints, as parameterized via $m$.  Since $\Delta z_s$ and $m$ both affect all ratio measurements for a single source galaxy bin, their impacts on the lensing ratios are largely degenerate.  Consequently, even for multiple lens redshift bins, when both $\Delta z_s$ and $m$ are left completely free, no useful level of self-calibration can be achieved, and the cosmological constraints are significantly degraded.  However, for the projected priors on $m$ of $\sigma(m) = 0.001$, the impact of marginalizing over $m$ on the cosmological constraints is negligible, given the projected statistical error bars on the ratios.  Note that the cosmological constraints presented in Fig.~\ref{fig:future_compare} include marginalization over $m$ with the fiducial $\sigma(m) = 0.001$ prior.

\section{Conclusions}
\label{sec:discussion}

Using a combination of galaxy position measurements and galaxy lensing maps from DES, and CMB lensing measurements from SPT and \textit{Planck}, we have measured several cosmological lensing ratios.  These ratios have the attractive feature that they can be modeled using only geometrical information (i.e. distances as a function of redshift), and do not depend on the galaxy-matter power spectrum.  Although lensing ratios use the CMB as a source plane, they are completely independent of the physics of baryon acoustic oscillations in the primordial plasma, making them a useful cross-check of geometrical constraints from the CMB and the BAO feature in the galaxy distribution.  Similarly, lensing ratios provide a test of cosmological distances that is completely independent of constraints from supernovae.   

Enabled by the well-understood photometric redshifts of the redMaGic galaxies, we have for the first time measured lensing ratios without the use of spectroscopic galaxy samples.  Each lensing ratio is constrained to 13 to 23\% precision, and the combined constraint from all ratios is roughly 10\%.  Using these measurements, we place constraints on curved $\Lambda$CDM cosmological models, finding consistency with the concordance cosmological model.  Our most interesting cosmological constraint is on $\Omega_{ k}$ and is shown in Fig.~\ref{fig:data_constraints}.  

We have also predicted the constraining power on lensing ratios of future experiments.  While previous forecasts have focused on spectroscopic identification of tracer galaxies, we argue that photometrically identified galaxies can be used, provided their redshifts can be constrained with redMaGiC-like accuracy.  Given this observation, we argue that the combination of data from LSST and CMB-S4 experiments will provide tight constraints on lensing ratios, achieving roughly 1.5\% precision for tracers distributed over $z \in [0.2, 0.7]$. Additionally, we showed that systematic uncertainty in the redshift estimates for the source galaxies significantly degrades the cosmological constraints from lensing ratios.  However, given the expected priors on the source galaxy redshift biases, the degradation from the source redshift uncertainty will be smaller than the statistical uncertainties. Moreover, we have found that using multiple lens and source bins allows for some self-calibration of the photometric redshifts, but not to the level of the expected priors. We have ignored the complication that photometric redshift errors may not be adequately parameterized by a single shift parameters as in Eq.~(\ref{eq:photozbias}).  Exploring the consequences of more generic redshift bias models is one avenue for future work.  We have also found that multiplicative shear biases will not be a limiting factor for lensing ratios given the expected priors on these parameters. 

When combined with geometrical constraints from the CMB, the lensing ratios explored in this work offer the possibility of deriving purely geometric constraints on the curvature of the Universe and the equation of state parameter of dark energy.  Analyses with future data sets will be able to significantly improve on current lensing ratio measurements, as seen in Table~\ref{table:error_surveys} and Fig.~\ref{fig:future_compare}.  While such future constraints would be interesting in their own right, their geometric nature also means that comparisons to cosmological probes that use growth and power spectrum information are particularly interesting.  Modified gravity, for instance, is expected to lead to differences in cosmological models inferred from geometry and growth measurements \citep[e.g.][]{Ruiz:2015}.  Exploring these possibilities with lensing ratios is another exciting avenue for future work.   

Part of the appeal of lensing ratios is their simplicity: they do not require complicated modeling of the two-point functions that they depend on.  Unfortunately, this simplicity comes at the cost of reduced sensitivity to cosmological parameters.  While lensing ratios have already been used to provide competitive constraints on systematics parameters \citep[e.g.][]{Prat:2017}, competitive cosmological constraints with lensing ratios have yet to be demonstrated.  Still, the geometric nature of the constraints, the fact that they are independent of the physics of BAO, and the fact that their sensitivity spans a wide range of redshifts make lensing ratios worth exploring with future data.  Furthermore, assuming cosmologists continue to measure two-point functions between galaxy density and gravitational lensing, lensing ratio constraints on cosmology come essentially for free.  

\section*{Acknowledgments}

 We have made use of \texttt{ChainConsumer} \citep{Hinton2016} to analyze Markov chains and to visualize multidimensional posteriors. 

Funding for the DES Projects has been provided by the U.S. Department of Energy, the U.S. National Science Foundation, the Ministry of Science and Education of Spain, 
the Science and Technology Facilities Council of the United Kingdom, the Higher Education Funding Council for England, the National Center for Supercomputing 
Applications at the University of Illinois at Urbana-Champaign, the Kavli Institute of Cosmological Physics at the University of Chicago, 
the Center for Cosmology and Astro-Particle Physics at the Ohio State University,
the Mitchell Institute for Fundamental Physics and Astronomy at Texas A\&M University, Financiadora de Estudos e Projetos, 
Funda{\c c}{\~a}o Carlos Chagas Filho de Amparo {\`a} Pesquisa do Estado do Rio de Janeiro, Conselho Nacional de Desenvolvimento Cient{\'i}fico e Tecnol{\'o}gico and 
the Minist{\'e}rio da Ci{\^e}ncia, Tecnologia e Inova{\c c}{\~a}o, the Deutsche Forschungsgemeinschaft and the Collaborating Institutions in the Dark Energy Survey. 

The Collaborating Institutions are Argonne National Laboratory, the University of California at Santa Cruz, the University of Cambridge, Centro de Investigaciones Energ{\'e}ticas, 
Medioambientales y Tecnol{\'o}gicas-Madrid, the University of Chicago, University College London, the DES-Brazil Consortium, the University of Edinburgh, 
the Eidgen{\"o}ssische Technische Hochschule (ETH) Z{\"u}rich, 
Fermi National Accelerator Laboratory, the University of Illinois at Urbana-Champaign, the Institut de Ci{\`e}ncies de l'Espai (IEEC/CSIC), 
the Institut de F{\'i}sica d'Altes Energies, Lawrence Berkeley National Laboratory, the Ludwig-Maximilians Universit{\"a}t M{\"u}nchen and the associated Excellence Cluster Universe, 
the University of Michigan, the National Optical Astronomy Observatory, the University of Nottingham, The Ohio State University, the University of Pennsylvania, the University of Portsmouth, 
SLAC National Accelerator Laboratory, Stanford University, the University of Sussex, Texas A\&M University, and the OzDES Membership Consortium.

Based in part on observations at Cerro Tololo Inter-American Observatory, National Optical Astronomy Observatory, which is operated by the Association of 
Universities for Research in Astronomy (AURA) under a cooperative agreement with the National Science Foundation.

The DES data management system is supported by the National Science Foundation under Grant Numbers AST-1138766 and AST-1536171.
The DES participants from Spanish institutions are partially supported by MINECO under grants AYA2015-71825, ESP2015-66861, FPA2015-68048, SEV-2016-0588, SEV-2016-0597, and MDM-2015-0509, 
some of which include ERDF funds from the European Union. IFAE is partially funded by the CERCA program of the Generalitat de Catalunya.
Research leading to these results has received funding from the European Research
Council under the European Union's Seventh Framework Program (FP7/2007-2013) including ERC grant agreements 240672, 291329, and 306478.
We  acknowledge support from the Australian Research Council Centre of Excellence for All-sky Astrophysics (CAASTRO), through project number CE110001020, and the Brazilian Instituto Nacional de Ci\^encia
e Tecnologia (INCT) e-Universe (CNPq grant 465376/2014-2).

This manuscript has been authored by Fermi Research Alliance, LLC under Contract No. DE-AC02-07CH11359 with the U.S. Department of Energy, Office of Science, Office of High Energy Physics. The United States Government retains and the publisher, by accepting the article for publication, acknowledges that the United States Government retains a non-exclusive, paid-up, irrevocable, world-wide license to publish or reproduce the published form of this manuscript, or allow others to do so, for United States Government purposes.

\bibliographystyle{mnras}
\bibliography{library.bib}

\begin{thebibliography}{}
\makeatletter
\relax
\def\mn@urlcharsother{\let\do\@makeother \do\$\do\&\do\#\do\^\do\_\do\%\do\~}
\def\mn@doi{\begingroup\mn@urlcharsother \@ifnextchar [ {\mn@doi@}
  {\mn@doi@[]}}
\def\mn@doi@[#1]#2{\def\@tempa{#1}\ifx\@tempa\@empty \href
  {http://dx.doi.org/#2} {doi:#2}\else \href {http://dx.doi.org/#2} {#1}\fi
  \endgroup}
\def\mn@eprint#1#2{\mn@eprint@#1:#2::\@nil}
\def\mn@eprint@arXiv#1{\href {http://arxiv.org/abs/#1} {{\tt arXiv:#1}}}
\def\mn@eprint@dblp#1{\href {http://dblp.uni-trier.de/rec/bibtex/#1.xml}
  {dblp:#1}}
\def\mn@eprint@#1:#2:#3:#4\@nil{\def\@tempa {#1}\def\@tempb {#2}\def\@tempc
  {#3}\ifx \@tempc \@empty \let \@tempc \@tempb \let \@tempb \@tempa \fi \ifx
  \@tempb \@empty \def\@tempb {arXiv}\fi \@ifundefined
  {mn@eprint@\@tempb}{\@tempb:\@tempc}{\expandafter \expandafter \csname
  mn@eprint@\@tempb\endcsname \expandafter{\@tempc}}}

\bibitem[\protect\citeauthoryear{{Abazajian} et~al.,}{{Abazajian}
  et~al.}{2016}]{CMBS4}
{Abazajian} K.~N.,  et~al., 2016, preprint, \href
  {http://adsabs.harvard.edu/abs/2016arXiv161002743A} {} (\mn@eprint {arXiv}
  {1610.02743})

\bibitem[\protect\citeauthoryear{{Abbott} et~al.,}{{Abbott}
  et~al.}{2018a}]{5x2pty1results}
{Abbott} T.~M.~C.,  et~al., 2018a, preprint, \href
  {http://adsabs.harvard.edu/abs/2018arXiv181002322A} {} (\mn@eprint {arXiv}
  {1810.02322})

\bibitem[\protect\citeauthoryear{Abbott et~al.}{Abbott
  et~al.}{2018b}]{DESy1:2017}
Abbott T. M.~C.,  et~al., 2018b, \mn@doi [Phys. Rev. D]
  {10.1103/PhysRevD.98.043526}, 98, 043526

\bibitem[\protect\citeauthoryear{Ade et~al.}{Ade
  et~al.}{2016}]{PlanckCollaboration2015}
Ade P. A.~R.,  et~al., 2016, \mn@doi [Astron. Astrophys.]
  {10.1051/0004-6361/201525830}, 594, A13

\bibitem[\protect\citeauthoryear{{Astropy Collaboration} et~al.,}{{Astropy
  Collaboration} et~al.}{2018}]{astropy2018}
{Astropy Collaboration} et~al., 2018, \mn@doi [\aj] {10.3847/1538-3881/aabc4f},
  \href {http://adsabs.harvard.edu/abs/2018AJ....156..123A} {156, 123}

\bibitem[\protect\citeauthoryear{{Aubourg} et~al.,}{{Aubourg}
  et~al.}{2015}]{BOSS:2015}
{Aubourg} {\'E}.,  et~al., 2015, \mn@doi [\prd] {10.1103/PhysRevD.92.123516},
  \href {http://adsabs.harvard.edu/abs/2015PhRvD..92l3516A} {92, 123516}

\bibitem[\protect\citeauthoryear{Baxter et~al.}{Baxter
  et~al.}{2016}]{Baxter2016}
Baxter E.,  et~al., 2016, \mn@doi [Mon. Not. R. Astron. Soc.]
  {10.1093/mnras/stw1584}, 461, 4099

\bibitem[\protect\citeauthoryear{{Baxter} et~al.,}{{Baxter}
  et~al.}{2019}]{Baxter:2018}
{Baxter} E.~J.,  et~al., 2019, \mn@doi [\prd] {10.1103/PhysRevD.99.023508},
  \href {http://adsabs.harvard.edu/abs/2019PhRvD..99b3508B} {99, 023508}

\bibitem[\protect\citeauthoryear{Benitez}{Benitez}{2000}]{Benitez2000a}
Benitez N.,  2000, \mn@doi [Astrophys. J.] {10.1086/308947}, 536, 571

\bibitem[\protect\citeauthoryear{{Benson}, {Cole}, {Frenk}, {Baugh}  \&
  {Lacey}}{{Benson} et~al.}{2000}]{Benson:2000}
{Benson} A.~J.,  {Cole} S.,  {Frenk} C.~S.,  {Baugh} C.~M.,   {Lacey} C.~G.,
  2000, \mn@doi [\mnras] {10.1046/j.1365-8711.2000.03101.x}, \href
  {http://adsabs.harvard.edu/abs/2000MNRAS.311..793B} {311, 793}

\bibitem[\protect\citeauthoryear{{Benson} et~al.,}{{Benson}
  et~al.}{2014}]{Benson:2014}
{Benson} B.~A.,  et~al., 2014, in Millimeter, Submillimeter, and Far-Infrared
  Detectors and Instrumentation for Astronomy VII. p. 91531P (\mn@eprint
  {arXiv} {1407.2973}), \mn@doi{10.1117/12.2057305}

\bibitem[\protect\citeauthoryear{{Bernstein}}{{Bernstein}}{2006}]{Bernstein2006}
{Bernstein} G.,  2006, \mn@doi [\apj] {10.1086/498079}, \href
  {http://adsabs.harvard.edu/abs/2006ApJ...637..598B} {637, 598}

\bibitem[\protect\citeauthoryear{{Bianchini} et~al.,}{{Bianchini}
  et~al.}{2015}]{Bianchini:2015}
{Bianchini} F.,  et~al., 2015, \mn@doi [\apj] {10.1088/0004-637X/802/1/64},
  \href {http://adsabs.harvard.edu/abs/2015ApJ...802...64B} {802, 64}

\bibitem[\protect\citeauthoryear{{Blazek}, {Mandelbaum}, {Seljak}  \&
  {Nakajima}}{{Blazek} et~al.}{2012}]{Blazek2012}
{Blazek} J.,  {Mandelbaum} R.,  {Seljak} U.,   {Nakajima} R.,  2012, \mn@doi
  [\jcap] {10.1088/1475-7516/2012/05/041}, \href
  {http://adsabs.harvard.edu/abs/2012JCAP...05..041B} {5, 041}

\bibitem[\protect\citeauthoryear{{Castro}, {Heavens}  \& {Kitching}}{{Castro}
  et~al.}{2005}]{Castro2005}
{Castro} P.~G.,  {Heavens} A.~F.,   {Kitching} T.~D.,  2005, \mn@doi [\prd]
  {10.1103/PhysRevD.72.023516}, \href
  {http://adsabs.harvard.edu/abs/2005PhRvD..72b3516C} {72, 023516}

\bibitem[\protect\citeauthoryear{{Cavaliere} \& {Fusco-Femiano}}{{Cavaliere} \&
  {Fusco-Femiano}}{1976}]{Cavaliere:1976}
{Cavaliere} A.,  {Fusco-Femiano} R.,  1976, \aap, \href
  {http://adsabs.harvard.edu/abs/1976A%26A....49..137C} {49, 137}

\bibitem[\protect\citeauthoryear{{Cawthon} et~al.,}{{Cawthon}
  et~al.}{2018}]{redmagicpz}
{Cawthon} R.,  et~al., 2018, \mn@doi [\mnras] {10.1093/mnras/sty2424}, \href
  {http://adsabs.harvard.edu/abs/2018MNRAS.481.2427C} {481, 2427}

\bibitem[\protect\citeauthoryear{{Chang} et~al.,}{{Chang}
  et~al.}{2018}]{des_mm_2017}
{Chang} C.,  et~al., 2018, \mn@doi [\mnras] {10.1093/mnras/stx3363}, \href
  {http://adsabs.harvard.edu/abs/2018MNRAS.475.3165C} {475, 3165}

\bibitem[\protect\citeauthoryear{{DES Collaboration}}{{DES
  Collaboration}}{2016}]{DES2016}
{DES Collaboration} 2016, \mn@doi [Mon. Not. R. Astron. Soc.]
  {10.1093/mnras/stw641}, 460, 1270

\bibitem[\protect\citeauthoryear{{DESI Collaboration} et~al.,}{{DESI
  Collaboration} et~al.}{2016}]{DESI}
{DESI Collaboration} et~al., 2016, preprint, \href
  {http://adsabs.harvard.edu/abs/2016arXiv161100036D} {} (\mn@eprint {arXiv}
  {1611.00036})

\bibitem[\protect\citeauthoryear{{Das} \& {Spergel}}{{Das} \&
  {Spergel}}{2009}]{Das:2009}
{Das} S.,  {Spergel} D.~N.,  2009, \mn@doi [\prd] {10.1103/PhysRevD.79.043509},
  \href {http://adsabs.harvard.edu/abs/2009PhRvD..79d3509D} {79, 043509}

\bibitem[\protect\citeauthoryear{{Davis} et~al.,}{{Davis}
  et~al.}{2017}]{Davis2017}
{Davis} C.,  et~al., 2017, preprint, \href
  {http://adsabs.harvard.edu/abs/2017arXiv171002517D} {} (\mn@eprint {arXiv}
  {1710.02517})

\bibitem[\protect\citeauthoryear{{Drlica-Wagner} et~al.}{{Drlica-Wagner}
  et~al.}{2018}]{y1gold}
{Drlica-Wagner} A.,  et~al., 2018, \mn@doi [Astrophys. J.]
  {10.3847/1538-4365/aab4f5}, \href
  {http://adsabs.harvard.edu/abs/2018ApJS..235...33D} {235, 33}

\bibitem[\protect\citeauthoryear{{Efstathiou} \& {Bond}}{{Efstathiou} \&
  {Bond}}{1999}]{Efstathiou:1999}
{Efstathiou} G.,  {Bond} J.~R.,  1999, \mn@doi [\mnras]
  {10.1046/j.1365-8711.1999.02274.x}, \href
  {http://adsabs.harvard.edu/abs/1999MNRAS.304...75E} {304, 75}

\bibitem[\protect\citeauthoryear{Elvin-Poole et~al.}{Elvin-Poole
  et~al.}{2018}]{wthetapaper}
Elvin-Poole J.,  et~al., 2018, \mn@doi [Phys. Rev. D]
  {10.1103/PhysRevD.98.042006}, 98, 042006

\bibitem[\protect\citeauthoryear{Flaugher et~al.}{Flaugher
  et~al.}{2015}]{Flaugher2015}
Flaugher B.,  et~al., 2015, \mn@doi [Astron. J.] {10.1088/0004-6256/150/5/150},
  150, 150

\bibitem[\protect\citeauthoryear{{Foreman-Mackey}, {Hogg}, {Lang}  \&
  {Goodman}}{{Foreman-Mackey} et~al.}{2013}]{emcee}
{Foreman-Mackey} D.,  {Hogg} D.~W.,  {Lang} D.,   {Goodman} J.,  2013, \mn@doi
  [PASP] {10.1086/670067}, 125, 306

\bibitem[\protect\citeauthoryear{{Gatti} et~al.,}{{Gatti}
  et~al.}{2018}]{Gatti2018}
{Gatti} M.,  et~al., 2018, \mn@doi [\mnras] {10.1093/mnras/sty466}, \href
  {http://adsabs.harvard.edu/abs/2018MNRAS.477.1664G} {477, 1664}

\bibitem[\protect\citeauthoryear{{Giannantonio} et~al.,}{{Giannantonio}
  et~al.}{2016}]{Giannantonio:2016}
{Giannantonio} T.,  et~al., 2016, \mn@doi [\mnras] {10.1093/mnras/stv2678},
  \href {http://adsabs.harvard.edu/abs/2016MNRAS.456.3213G} {456, 3213}

\bibitem[\protect\citeauthoryear{Hartlap, Simon  \& Schneider}{Hartlap
  et~al.}{2007}]{Hartlap2007}
Hartlap J.,  Simon P.,   Schneider P.,  2007, \mn@doi [Astron. Astrophys.]
  {10.1051/0004-6361:20066170}, 464, 399

\bibitem[\protect\citeauthoryear{{Heavens}}{{Heavens}}{2003}]{Heavens2003}
{Heavens} A.,  2003, \mn@doi [\mnras] {10.1046/j.1365-8711.2003.06780.x}, \href
  {http://adsabs.harvard.edu/abs/2003MNRAS.343.1327H} {343, 1327}

\bibitem[\protect\citeauthoryear{{Heavens}, {Kitching}  \& {Taylor}}{{Heavens}
  et~al.}{2006}]{Heavens2006}
{Heavens} A.~F.,  {Kitching} T.~D.,   {Taylor} A.~N.,  2006, \mn@doi [\mnras]
  {10.1111/j.1365-2966.2006.11006.x}, \href
  {http://adsabs.harvard.edu/abs/2006MNRAS.373..105H} {373, 105}

\bibitem[\protect\citeauthoryear{{Henderson} et~al.,}{{Henderson}
  et~al.}{2016}]{Henderson:2016}
{Henderson} S.~W.,  et~al., 2016, \mn@doi [Journal of Low Temperature Physics]
  {10.1007/s10909-016-1575-z}, \href
  {http://adsabs.harvard.edu/abs/2016JLTP..184..772H} {184, 772}

\bibitem[\protect\citeauthoryear{{Hinton}}{{Hinton}}{2016}]{Hinton2016}
{Hinton} S.~R.,  2016, \mn@doi [The Journal of Open Source Software]
  {10.21105/joss.00045}, \href
  {http://adsabs.harvard.edu/abs/2016JOSS....1...45H} {1, 00045}

\bibitem[\protect\citeauthoryear{{Hoyle} et~al.,}{{Hoyle}
  et~al.}{2018}]{Hoyle2018}
{Hoyle} B.,  et~al., 2018, \mn@doi [\mnras] {10.1093/mnras/sty957}, \href
  {http://adsabs.harvard.edu/abs/2018MNRAS.478..592H} {478, 592}

\bibitem[\protect\citeauthoryear{{Hu} \& {Okamoto}}{{Hu} \&
  {Okamoto}}{2002}]{Hu:2002}
{Hu} W.,  {Okamoto} T.,  2002, \mn@doi [\apj] {10.1086/341110}, \href
  {http://adsabs.harvard.edu/abs/2002ApJ...574..566H} {574, 566}

\bibitem[\protect\citeauthoryear{{Hu}, {Holz}  \& {Vale}}{{Hu}
  et~al.}{2007}]{Hu:2007}
{Hu} W.,  {Holz} D.~E.,   {Vale} C.,  2007, \mn@doi [\prd]
  {10.1103/PhysRevD.76.127301}, \href
  {http://adsabs.harvard.edu/abs/2007PhRvD..76l7301H} {76, 127301}

\bibitem[\protect\citeauthoryear{Huff \& Mandelbaum}{Huff \&
  Mandelbaum}{2017}]{Huff2017}
Huff E.,  Mandelbaum R.,  2017, preprint (\mn@eprint {arXiv} {1702.02600})

\bibitem[\protect\citeauthoryear{{Jain} \& {Taylor}}{{Jain} \&
  {Taylor}}{2003}]{Jain:2003}
{Jain} B.,  {Taylor} A.,  2003, \mn@doi [Physical Review Letters]
  {10.1103/PhysRevLett.91.141302}, \href
  {http://adsabs.harvard.edu/abs/2003PhRvL..91n1302J} {91, 141302}

\bibitem[\protect\citeauthoryear{Jarvis, Bernstein  \& Jain}{Jarvis
  et~al.}{2004}]{Jarvis2004}
Jarvis M.,  Bernstein G.,   Jain B.,  2004, \mn@doi [Mon. Not. R. Astron. Soc.]
  {10.1111/j.1365-2966.2004.07926.x}, 352, 338

\bibitem[\protect\citeauthoryear{{Kaiser} \& {Squires}}{{Kaiser} \&
  {Squires}}{1993}]{Kaiser1993}
{Kaiser} N.,  {Squires} G.,  1993, \mn@doi [Astrophys. J.] {10.1086/172297},
  \href {http://adsabs.harvard.edu/abs/1993ApJ...404..441K} {404, 441}

\bibitem[\protect\citeauthoryear{{Kitching} et~al.,}{{Kitching}
  et~al.}{2015}]{Kitching2015}
{Kitching} T.~D.,  et~al., 2015, preprint, \href
  {http://adsabs.harvard.edu/abs/2015arXiv151203627K} {} (\mn@eprint {arXiv}
  {1512.03627})

\bibitem[\protect\citeauthoryear{{Krause} et~al.,}{{Krause}
  et~al.}{2017}]{Krause:2017}
{Krause} E.,  et~al., 2017, preprint, \href
  {http://adsabs.harvard.edu/abs/2017arXiv170609359K} {} (\mn@eprint {arXiv}
  {1706.09359})

\bibitem[\protect\citeauthoryear{{LSST Dark Energy Science
  Collaboration}}{{LSST Dark Energy Science Collaboration}}{2012}]{LSST}
{LSST Dark Energy Science Collaboration} 2012, preprint, \href
  {http://adsabs.harvard.edu/abs/2012arXiv1211.0310L} {} (\mn@eprint {arXiv}
  {1211.0310})

\bibitem[\protect\citeauthoryear{{LSST Science Collaboration} et~al.,}{{LSST
  Science Collaboration} et~al.}{2009}]{LSST:ScienceBook}
{LSST Science Collaboration} et~al., 2009, preprint, \href
  {http://adsabs.harvard.edu/abs/2009arXiv0912.0201L} {} (\mn@eprint {arXiv}
  {0912.0201})

\bibitem[\protect\citeauthoryear{{Leistedt}, {McEwen}, {B{\"u}ttner}  \&
  {Peiris}}{{Leistedt} et~al.}{2017}]{Leistedt2017}
{Leistedt} B.,  {McEwen} J.~D.,  {B{\"u}ttner} M.,   {Peiris} H.~V.,  2017,
  \mn@doi [\mnras] {10.1093/mnras/stw3176}, \href
  {http://adsabs.harvard.edu/abs/2017MNRAS.466.3728L} {466, 3728}

\bibitem[\protect\citeauthoryear{Madhavacheril \& Hill}{Madhavacheril \&
  Hill}{2018}]{Madhavacheril:2018}
Madhavacheril M.~S.,  Hill J.~C.,  2018, \mn@doi [Phys. Rev. D]
  {10.1103/PhysRevD.98.023534}, 98, 023534

\bibitem[\protect\citeauthoryear{{McClintock} et~al.,}{{McClintock}
  et~al.}{2019}]{McClintock:2018}
{McClintock} T.,  et~al., 2019, \mn@doi [\mnras] {10.1093/mnras/sty2711}, \href
  {http://adsabs.harvard.edu/abs/2019MNRAS.482.1352M} {482, 1352}

\bibitem[\protect\citeauthoryear{{Melchior} et~al.,}{{Melchior}
  et~al.}{2017}]{Melchior:2017}
{Melchior} P.,  et~al., 2017, \mn@doi [\mnras] {10.1093/mnras/stx1053}, \href
  {http://adsabs.harvard.edu/abs/2017MNRAS.469.4899M} {469, 4899}

\bibitem[\protect\citeauthoryear{{Miyatake}, {Madhavacheril}, {Sehgal},
  {Slosar}, {Spergel}, {Sherwin}  \& {van Engelen}}{{Miyatake}
  et~al.}{2017}]{Miyatake:2017}
{Miyatake} H.,  {Madhavacheril} M.~S.,  {Sehgal} N.,  {Slosar} A.,  {Spergel}
  D.~N.,  {Sherwin} B.,   {van Engelen} A.,  2017, \mn@doi [Physical Review
  Letters] {10.1103/PhysRevLett.118.161301}, \href
  {http://adsabs.harvard.edu/abs/2017PhRvL.118p1301M} {118, 161301}

\bibitem[\protect\citeauthoryear{{Omori} et~al.,}{{Omori}
  et~al.}{2017}]{Omori2017}
{Omori} Y.,  et~al., 2017, \mn@doi [Astrophys. J.] {10.3847/1538-4357/aa8d1d},
  \href {http://adsabs.harvard.edu/abs/2017ApJ...849..124O} {849, 124}

\bibitem[\protect\citeauthoryear{{Omori} et~al.,}{{Omori}
  et~al.}{2018}]{NKpaper}
{Omori} Y.,  et~al., 2018, preprint, \href
  {http://adsabs.harvard.edu/abs/2018arXiv181002342O} {} (\mn@eprint {arXiv}
  {1810.02342})

\bibitem[\protect\citeauthoryear{Prat et~al.,}{Prat et~al.}{2018}]{Prat:2017}
Prat J.,  et~al., 2018, \mn@doi [Phys. Rev. D] {10.1103/PhysRevD.98.042005},
  98, 042005

\bibitem[\protect\citeauthoryear{{Rozo} et~al.}{{Rozo} et~al.}{2016}]{Rozo2015}
{Rozo} E.,  et~al., 2016, \mn@doi [Mon. Not. R. Astron. Soc.]
  {10.1093/mnras/stw1281}, \href
  {http://adsabs.harvard.edu/abs/2016MNRAS.461.1431R} {461, 1431}

\bibitem[\protect\citeauthoryear{{Ruiz} \& {Huterer}}{{Ruiz} \&
  {Huterer}}{2015}]{Ruiz:2015}
{Ruiz} E.~J.,  {Huterer} D.,  2015, \mn@doi [\prd]
  {10.1103/PhysRevD.91.063009}, \href
  {http://adsabs.harvard.edu/abs/2015PhRvD..91f3009R} {91, 063009}

\bibitem[\protect\citeauthoryear{{Samuroff} et~al.,}{{Samuroff}
  et~al.}{2018}]{y1-neighbours}
{Samuroff} S.,  et~al., 2018, \mn@doi [\mnras] {10.1093/mnras/stx3282}, \href
  {http://adsabs.harvard.edu/abs/2018MNRAS.475.4524S} {475, 4524}

\bibitem[\protect\citeauthoryear{{Schaan}, {Krause}, {Eifler}, {Dor{\'e}},
  {Miyatake}, {Rhodes}  \& {Spergel}}{{Schaan} et~al.}{2017}]{Schaan:2017}
{Schaan} E.,  {Krause} E.,  {Eifler} T.,  {Dor{\'e}} O.,  {Miyatake} H.,
  {Rhodes} J.,   {Spergel} D.~N.,  2017, \mn@doi [\prd]
  {10.1103/PhysRevD.95.123512}, \href
  {http://adsabs.harvard.edu/abs/2017PhRvD..95l3512S} {95, 123512}

\bibitem[\protect\citeauthoryear{{Schneider}}{{Schneider}}{1996}]{Schneider1996}
{Schneider} P.,  1996, \mn@doi [Mon. Not. R. Astron. Soc.]
  {10.1093/mnras/283.3.837}, \href
  {http://adsabs.harvard.edu/abs/1996MNRAS.283..837S} {283, 837}

\bibitem[\protect\citeauthoryear{Sheldon \& Huff}{Sheldon \&
  Huff}{2017}]{Sheldon2017}
Sheldon E.~S.,  Huff E.~M.,  2017, \mn@doi [Astrophys. J.]
  {10.3847/1538-4357/aa704b}, 841, 24

\bibitem[\protect\citeauthoryear{{Takada} \& {Hu}}{{Takada} \&
  {Hu}}{2013}]{Takada:2013}
{Takada} M.,  {Hu} W.,  2013, \mn@doi [\prd] {10.1103/PhysRevD.87.123504},
  \href {https://ui.adsabs.harvard.edu/\#abs/2013PhRvD..87l3504T} {87, 123504}

\bibitem[\protect\citeauthoryear{{Takahashi}, {Sato}, {Nishimichi}, {Taruya}
  \& {Oguri}}{{Takahashi} et~al.}{2012}]{Takahashi:2012}
{Takahashi} R.,  {Sato} M.,  {Nishimichi} T.,  {Taruya} A.,   {Oguri} M.,
  2012, \mn@doi [\apj] {10.1088/0004-637X/761/2/152}, \href
  {http://adsabs.harvard.edu/abs/2012ApJ...761..152T} {761, 152}

\bibitem[\protect\citeauthoryear{Troxel et~al.}{Troxel
  et~al.}{2018}]{shearcorr}
Troxel M.~A.,  et~al., 2018, \mn@doi [Phys. Rev. D]
  {10.1103/PhysRevD.98.043528}, 98, 043528

\bibitem[\protect\citeauthoryear{{Wallis}, {McEwen}, {Kitching}, {Leistedt}  \&
  {Plouviez}}{{Wallis} et~al.}{2017}]{Wallis2017}
{Wallis} C.~G.~R.,  {McEwen} J.~D.,  {Kitching} T.~D.,  {Leistedt} B.,
  {Plouviez} A.,  2017, preprint, \href
  {http://adsabs.harvard.edu/abs/2017arXiv170309233W} {} (\mn@eprint {arXiv}
  {1703.09233})

\bibitem[\protect\citeauthoryear{{White}, {Song}  \& {Percival}}{{White}
  et~al.}{2009}]{White:2009}
{White} M.,  {Song} Y.-S.,   {Percival} W.~J.,  2009, \mn@doi [\mnras]
  {10.1111/j.1365-2966.2008.14379.x}, \href
  {http://adsabs.harvard.edu/abs/2009MNRAS.397.1348W} {397, 1348}

\bibitem[\protect\citeauthoryear{{Zuntz}, {Sheldon}  et~al.}{{Zuntz}
  et~al.}{2018}]{shearcat}
{Zuntz} J.,  {Sheldon} E.,   et~al., 2018, \mn@doi [Mon. Not. R. Astron. Soc.]
  {10.1093/mnras/sty2219}, \href
  {http://adsabs.harvard.edu/abs/2018MNRAS.tmp.2129Z} {}

\bibitem[\protect\citeauthoryear{{van Daalen}, {Schaye}, {Booth}  \& {Dalla
  Vecchia}}{{van Daalen} et~al.}{2011}]{vanDaalen:2011}
{van Daalen} M.~P.,  {Schaye} J.,  {Booth} C.~M.,   {Dalla Vecchia} C.,  2011,
  \mn@doi [\mnras] {10.1111/j.1365-2966.2011.18981.x}, \href
  {http://adsabs.harvard.edu/abs/2011MNRAS.415.3649V} {415, 3649}

\makeatother
\end{thebibliography}

\section{Affiliations}
\textit{
$^{1}$ Institut de F\'{\i}sica d'Altes Energies (IFAE), The Barcelona Institute of Science and Technology, Campus UAB, 08193 Bellaterra (Barcelona) Spain\\
$^{2}$ Department of Physics and Astronomy, University of Pennsylvania, Philadelphia, PA 19104, USA\\
$^{3}$ Kavli Institute for Cosmological Physics, University of Chicago, Chicago, IL 60637, USA\\
$^{4}$ Instituci\'o Catalana de Recerca i Estudis Avan\c{c}ats, E-08010 Barcelona, Spain\\
$^{5}$ Institut d'Estudis Espacials de Catalunya (IEEC), 08193 Barcelona, Spain\\
$^{6}$ Institute of Space Sciences (ICE, CSIC),  Campus UAB, Carrer de Can Magrans, s/n,  08193 Barcelona, Spain\\
$^{7}$ Institute of Cosmology \& Gravitation, University of Portsmouth, Portsmouth, PO1 3FX, UK\\
$^{8}$ Department of Astronomy and Astrophysics, University of Chicago, Chicago, IL, 60637, USA\\
$^{9}$ Kavli Institute for Particle Astrophysics \& Cosmology, P. O. Box 2450, Stanford University, Stanford, CA 94305, USA\\
$^{10}$ Centro de Investigaciones Energ\'eticas, Medioambientales y Tecnol\'ogicas (CIEMAT), Madrid, Spain\\
$^{11}$ Department of Physics, Carnegie Mellon University, Pittsburgh, Pennsylvania 15312, USA\\
$^{12}$ Department of Astronomy/Steward Observatory, 933 North Cherry Avenue, Tucson, AZ 85721-0065, USA\\
$^{13}$ Jet Propulsion Laboratory, California Institute of Technology, 4800 Oak Grove Dr., Pasadena, CA 91109, USA\\
$^{14}$ Max Planck Institute for Extraterrestrial Physics, Giessenbachstrasse, 85748 Garching, Germany\\
$^{15}$ Universit\"ats-Sternwarte, Fakult\"at f\"ur Physik, Ludwig-Maximilians Universit\"at M\"unchen, Scheinerstr. 1, 81679 M\"unchen, Germany\\
$^{16}$ Einstein Fellow\\
$^{17}$ SLAC National Accelerator Laboratory, Menlo Park, CA 94025, USA\\
$^{18}$ Department of Physics \& Astronomy, University College London, Gower Street, London, WC1E 6BT, UK\\
$^{19}$ Department of Physics, ETH Zurich, Wolfgang-Pauli-Strasse 16, CH-8093 Zurich, Switzerland\\
$^{20}$ Department of Physics and McGill Space Institute, McGill University, Montreal, Quebec H3A 2T8, Canada\\
$^{21}$ Canadian Institute for Advanced Research, CIFAR Program in Gravity and the Extreme Universe, Toronto, ON, M5G 1Z8, Canada\\
$^{22}$ Department of Astronomy, University of Illinois at Urbana-Champaign, 1002 W. Green Street, Urbana, IL 61801, USA\\
$^{23}$ Department of Physics, University of Illinois Urbana-Champaign, 1110 W. Green Street, Urbana, IL 61801, USA\\
$^{24}$ Center for Cosmology and Astro-Particle Physics, The Ohio State University, Columbus, OH 43210, USA\\
$^{25}$ Department of Physics, The Ohio State University, Columbus, OH 43210, USA\\
$^{26}$ School of Physics, University of Melbourne, Parkville, VIC 3010, Australia\\
$^{27}$ Brookhaven National Laboratory, Bldg 510, Upton, NY 11973, USA\\
$^{28}$ Institute for Astronomy, University of Edinburgh, Edinburgh EH9 3HJ, UK\\
$^{29}$ Cerro Tololo Inter-American Observatory, National Optical Astronomy Observatory, Casilla 603, La Serena, Chile\\
$^{30}$ Department of Physics and Electronics, Rhodes University, PO Box 94, Grahamstown, 6140, South Africa\\
$^{31}$ Fermi National Accelerator Laboratory, P. O. Box 500, Batavia, IL 60510, USA\\
$^{32}$ Department of Physics, University of California, Davis, CA, USA 95616\\
$^{33}$ Department of Astronomy and Astrophysics, University of Chicago, Chicago, IL 60637, USA\\
$^{34}$ CNRS, UMR 7095, Institut d'Astrophysique de Paris, F-75014, Paris, France\\
$^{35}$ Sorbonne Universit\'es, UPMC Univ Paris 06, UMR 7095, Institut d'Astrophysique de Paris, F-75014, Paris, France\\
$^{36}$ High Energy Physics Division, Argonne National Laboratory, Argonne, IL, USA 60439\\
$^{37}$ Department of Physics, University of Chicago, Chicago, IL 60637, USA\\
$^{38}$ Enrico Fermi Institute, University of Chicago, Chicago, IL 60637, USA\\
$^{39}$ National Center for Supercomputing Applications, 1205 West Clark St., Urbana, IL 61801, USA\\
$^{40}$ California Institute of Technology, Pasadena, CA, USA 91125\\
$^{41}$ Laborat\'orio Interinstitucional de e-Astronomia - LIneA, Rua Gal. Jos\'e Cristino 77, Rio de Janeiro, RJ - 20921-400, Brazil\\
$^{42}$ Observat\'orio Nacional, Rua Gal. Jos\'e Cristino 77, Rio de Janeiro, RJ - 20921-400, Brazil\\
$^{43}$ Department of Physics, IIT Hyderabad, Kandi, Telangana 502285, India\\
$^{44}$ Excellence Cluster Universe, Boltzmannstr.\ 2, 85748 Garching, Germany\\
$^{45}$ Faculty of Physics, Ludwig-Maximilians-Universit\"at, Scheinerstr. 1, 81679 Munich, Germany\\
$^{46}$ Center for Astrophysics and Space Astronomy, Department of Astrophysical and Planetary Sciences, University of Colorado, Boulder, CO, 80309\\
$^{47}$ Department of Astronomy, University of Michigan, Ann Arbor, MI 48109, USA\\
$^{48}$ Department of Physics, University of Michigan, Ann Arbor, MI 48109, USA\\
$^{49}$ Instituto de Fisica Teorica UAM/CSIC, Universidad Autonoma de Madrid, 28049 Madrid, Spain\\
$^{50}$ Department of Physics, University of California, Berkeley, CA, USA 94720\\
$^{51}$ European Southern Observatory, Karl-Schwarzschild-Stra{\ss}e 2, 85748 Garching, Germany\\
$^{52}$ Institute of Astronomy, University of Cambridge, Madingley Road, Cambridge CB3 0HA, UK\\
$^{53}$ Kavli Institute for Cosmology, University of Cambridge, Madingley Road, Cambridge CB3 0HA, UK\\
$^{54}$ Physics Division, Lawrence Berkeley National Laboratory, Berkeley, CA, USA 94720\\
$^{55}$ Department of Physics, University of Colorado, Boulder, CO, 80309\\
$^{56}$ University of Chicago, Chicago, IL 60637, USA\\
$^{57}$ Harvard-Smithsonian Center for Astrophysics, Cambridge, MA 02138, USA\\
$^{58}$ Santa Cruz Institute for Particle Physics, Santa Cruz, CA 95064, USA\\
$^{59}$ Australian Astronomical Observatory, North Ryde, NSW 2113, Australia\\
$^{60}$ Departamento de F\'isica Matem\'atica, Instituto de F\'isica, Universidade de S\~ao Paulo, CP 66318, S\~ao Paulo, SP, 05314-970, Brazil\\
$^{61}$ Steward Observatory, University of Arizona, 933 North Cherry Avenue, Tucson, AZ 85721\\
$^{62}$ George P. and Cynthia Woods Mitchell Institute for Fundamental Physics and Astronomy, and Department of Physics and Astronomy, Texas A\&M University, College Station, TX 77843,  USA\\
$^{63}$ Department of Astrophysical Sciences, Princeton University, Peyton Hall, Princeton, NJ 08544, USA\\
$^{64}$ Department of Physics, University of Chicago, Chicago, IL, USA 6063\\
$^{65}$ Dunlap Institute for Astronomy \& Astrophysics, University of Toronto, 50 St George St, Toronto, ON, M5S 3H4, Canada\\
$^{66}$ Department of Physics, University of Minnesota, Minneapolis, MN, USA 55455\\
$^{67}$ Department of Physics and Astronomy, Pevensey Building, University of Sussex, Brighton, BN1 9QH, UK\\
$^{68}$ Physics Department, Center for Education and Research in Cosmology and Astrophysics, Case Western Reserve University,Cleveland, OH, USA 44106\\
$^{69}$ Liberal Arts Department, School of the Art Institute of Chicago, Chicago, IL, USA 60603\\
$^{70}$ School of Physics and Astronomy, University of Southampton,  Southampton, SO17 1BJ, UK\\
$^{71}$ Brandeis University, Physics Department, 415 South Street, Waltham MA 02453\\
$^{72}$ Instituto de F\'isica Gleb Wataghin, Universidade Estadual de Campinas, 13083-859, Campinas, SP, Brazil\\
$^{73}$ Dept. of Physics, Stanford University, 382 Via Pueblo Mall, Stanford, CA 94305\\
$^{74}$ Computer Science and Mathematics Division, Oak Ridge National Laboratory, Oak Ridge, TN 37831\\
$^{75}$ Department of Astronomy \& Astrophysics, University of Toronto, 50 St George St, Toronto, ON, M5S 3H4, Canada\\
$^{76}$ Argonne National Laboratory, 9700 South Cass Avenue, Lemont, IL 60439, USA\\
$^{77}$ Berkeley Center for Cosmological Physics, Department of Physics, University of California, and Lawrence Berkeley National Labs, Berkeley, CA, USA 94720\\
}

\appendix

\section{Test of Gaussian approximation to ratio posteriors}
\label{sec:gaussianity}
\begin{figure*}
\centering	\includegraphics[width=0.8\textwidth]{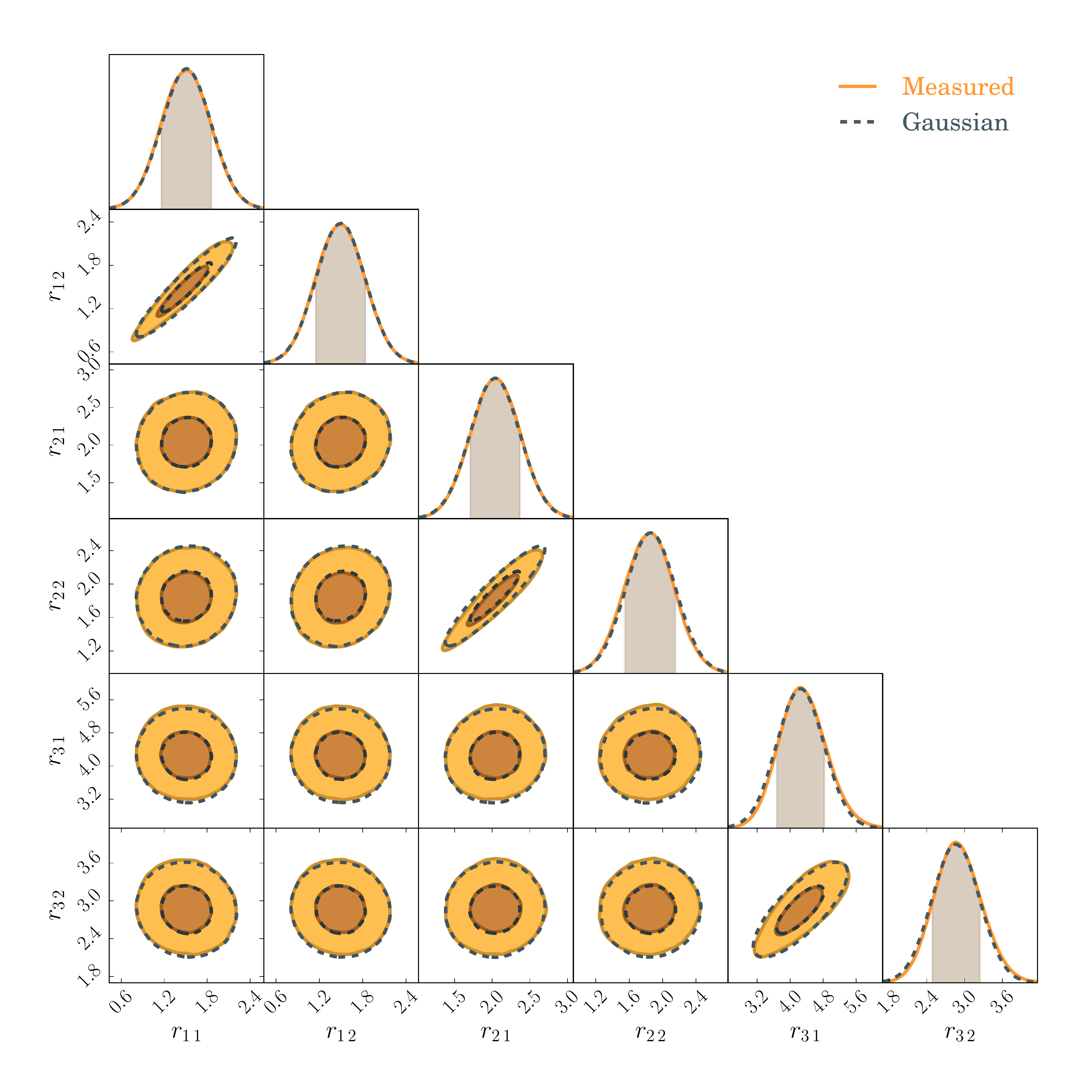}
    \caption{Measured posteriors on the lensing ratios compared to a multivariate Gaussian drawn from the measured covariance between the ratios centered in the same value. $r_{ij}$ is the ratio between the measurements in the CMB lensing map and the lens bin $i$ and the convergence map in source bin $j$ and same lens bin, as defined in Eq.~(\ref{eq:ratio_betas}).}
    \label{fig:ratio-posteriors}
\end{figure*}

In Fig.~\ref{fig:ratio-posteriors} we show the posteriors on the lensing ratios obtained from the fitting procedure to the two-point correlation functions using an MCMC, described in detail in Sec.~\ref{subsec:extracting_constraints_ratios}. These posteriors serve as the likelihood to then measure the cosmological parameters running a second MCMC, as can be seen in Eq.~(\ref{eq:posterior-cosmopars}). In our analysis, for simplicity, we assume this likelihood is a multivariate Gaussian with a covariance coming from the fitting procedure of Sec.~\ref{subsec:extracting_constraints_ratios}. We test this assumption in Fig.~\ref{fig:ratio-posteriors}, where we compare the measured lensing ratio posteriors with contours drawn from a multivariate Gaussian centered at the same value, and using the measured covariance, finding that they are indeed very similar. 

\bsp	
\label{lastpage}
\end{document}